\documentclass[aps,prb,twocolumn,superscriptaddress,showpacs]{revtex4-1}
\usepackage{graphicx}
\usepackage{multirow}

\begin{document}

\title{Effects of magnetic field and twin domains on magnetostructural phase mixture in Mn$_3$O$_4$: Raman scattering studies of untwinned crystals}

\author{T.~Byrum}\thanks{These coauthors contributed equally to this work.}
\author{S.~L.~Gleason}\thanks{These coauthors contributed equally to this work.}
\author{A.~Thaler}
\author{G.~J.~MacDougall}
\author{S.~L.~Cooper}
\affiliation{Department of Physics and Frederick Seitz Materials Research Laboratory, University of Illinois, Urbana, Illinois 61801, USA}

\date{\today}

\begin{abstract}

The ferrimagnetic spinel Mn$_3$O$_4$ exhibits large and anisotropic changes in electronic and structural properties in response to an applied magnetic field.
These changes are thought to result from the field-dependent tuning---via strong spin-lattice coupling---between two nearly degenerate magnetostructural phases.
Recent variable-magnetic-field studies of Mn$_3$O$_4$ have been performed on melt-grown crystals, which can exhibit twin domains due to a Jahn-Teller structural transition below the melting temperature.
Because of the near degeneracy of the magnetostructural phases, however, strain associated with the twin domains likely affects the magnetic responses of Mn$_3$O$_4$.
In this report, we present a variable-magnetic-field Raman scattering study of untwinned Mn$_3$O$_4$ crystals grown out of a flux below the Jahn-Teller structural transition.
We measure distinct $\mathbf{q}=0$ magnetic and vibrational excitation spectra for each isolated magnetostructural phase of untwinned Mn$_3$O$_4$ crystals and determine the symmetries of the observed excitations.
We determine how the magnetostructural phase mixture changes in response to magnetic fields applied in the magnetic easy plane.
Lastly, by comparing results on flux- and melt-grown Mn$_3$O$_4$ crystals, we show that the intrinsic mixture of the two magnetostructural phases is indeed strongly influenced by the presence of twin domains.

\end{abstract}


\maketitle

\section{\label{introduction}Introduction}

Strongly correlated materials are characterized by a competition among multiple important interactions involving spin, charge, lattice, and orbital degrees of freedom, which often leads to a near degeneracy between competing phases.\cite{Dagotto2005}
Significant consequences of this near degeneracy include the coexistence of phases and/or the `colossal' tunability of physical properties in response to external perturbations such as magnetic field and pressure.\cite{Dagotto2008, Ahn2013, Wu2003, Snow2001, Kivelson1998}
Examples of such tunable properties include multiferroic behavior,\cite{Kimura2003} colossal magnetoresistance,\cite{Ramirez1997} magnetically tuned shape-memory effects,\cite{Lavrov2002,Kim2010} and magnetothermal and magnetodielectric behavior.\cite{Lawes2003, Gschneidner2005, Tackett2007}
Clarifying the mechanisms connecting nearly degenerate phases and tunable behavior in specific materials has been one of the key challenges of condensed matter physics in recent years.

\begin{figure}[h]
\includegraphics[width=1.0\linewidth]{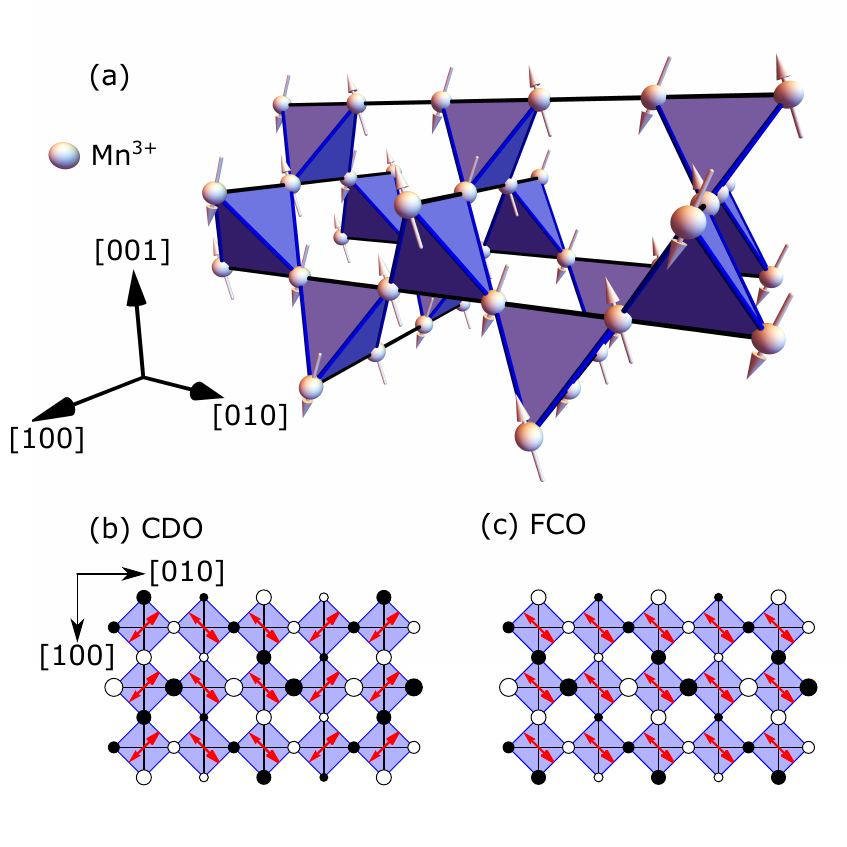}
\caption{\label{lattice} (Color online) 
(a)~Schematic view of the frustrating Mn$^{3+}$ sublattice, composed of corner-sharing tetrahedra. Antiferromagnetic chains along the $[100]$ and $[010]$ directions are indicated by black lines.
(b),(c) Proposed modulations of the Mn$^{3+}$ spins and Mn$^{3+}_4$ tetrahedra, given by Nii et al.\cite{Nii2013} and Chung et al.,\cite{Chung2013b} projected onto the $(001)$ plane for the (b) cell-doubled-orthorhombic (CDO) phase and the (c) face-centered-orthombic (FCO) phase. Black and white circles specify the c-axis components of the Mn$^{3+}$ spins and their relative sizes indicate position along the \textit{c}-axis. Red arrows indicate the local distortions of the Mn$^{3+}$ tetrahedra.}
\end{figure}

The spinel crystal structure (chemical formula $AB_2X_4$) is a commonly occurring structure among complex transition metal oxides and chalcogenides.\cite{Lacroix2011}
Many spinels display strong correlations between spin, charge, orbital, and lattice degrees of freedom due to a magnetically frustrating sublattice of corner-sharing tetrahedra (see Fig.~\ref{lattice}(a)), as well as the anisotropy and spatial extent of the $d$ orbitals.
Consequently, a wide variety of phenomena have been reported for spinels, including superconductivity,\cite{McCallum1976} charge ordering,\cite{Irizawa2011} heavy fermion behavior,\cite{Kopec2009} and multiferroicity.\cite{Yamasaki2006,Dey2014,Giovannetti2011,Maignan2012,Singh2011}

The ferrimagnetic spinel Mn$_3$O$_4$ ($A=\text{Mn}^{2+},B=\text{Mn}^{3+},X=\text{O}^{2-}$) also exhibits novel phenomena in the magnetically-ordered state, including large magnetodielectric and magnetoelastic responses.\cite{Tackett2007,Suzuki2008,Nii2013}
Magnetic-field-induced changes in the low-frequency dielectric constant and crystal dimensions have been attributed to field-induced transitions between two distinct magnetostructural phases, which are discussed below.\cite{Nii2013}
A high-resolution x-ray diffraction study recently revealed that these two magnetostructural phases in fact coexist below the N\'{e}el temperature in polycrystalline Mn$_3$O$_4$,\cite{Kemei2014c} demonstrating the near degeneracy of their free energies.
Thus, Mn$_3$O$_4$ belongs to the class of strongly correlated materials whose bulk physical parameters are `tunable' due to distinct phases existing in close proximity.

Mn$_3$O$_4$ is a tetragonally-distorted spinel at room temperature as a result of a Jahn-Teller cubic-to-tetragonal transition at $T=1440$~K.\cite{VanHook1958}
Mn$_3$O$_4$ first magnetically orders at $T=42$~K in a triangular configuration, with the spins lying approximately in the $(100)$ plane and with the net magnetization along the $[010]$ direction,\cite{Jensen1974,Chardon1986} where the crystallographic indices here and throughout the paper refer to the conventional (body-) centered-tetragonal setting.
The $c$-axis components of the Mn$^{3+}$ spins form antiferromagnetic chains along the $\langle 100 \rangle$ directions.\cite{Jensen1974}
Below $T=39$~K, an incommensurate magnetic modulation develops in half of the Mn$^{3+}$ spins, with a propagation vector that evolves with temperature until it becomes commensurate with the lattice below $T=34$~K, doubling the magnetic unit cell along the $[010]$ direction.\cite{Jensen1974,Chardon1986}

A structural modulation with a propagation vector identical to that of the magnetic modulation has also been observed,\cite{Nii2013} demonstrating the strong spin-lattice coupling in Mn$_3$O$_4$.
This strong coupling is proposed to originate from the geometric frustration of the Mn$^{3+}$ spins, which is only partially relieved by the high-temperature, cubic-to-tetragonal distortion.\cite{Nii2013,Chung2013b}
The resultant exchange striction in the magnetically-ordered state leads to a cell-doubled-orthorhombic (CDO) structure whose modulation is of the form pictured in Fig.~\ref{lattice}(b) below $T=34$~K.

It is known that structural and magnetic symmetry changes in Mn$_3$O$_4$ can be induced with modest magnetic fields applied in the $(001)$ plane (magnetic easy plane).\cite{Kim2010,Kim2011a,Nii2013}
For example, cooling a sample in the presence of a $H=1$~T magnetic field applied along the $[110]$ direction drives a uniform orthorhombic distortion of the lattice via spin-orbit coupling and inverse exchange striction.\cite{Nii2013}
The resultant face-centered-orthorhombic (FCO) structure is proposed to be of the form pictured in Fig.~\ref{lattice}(c).

Interestingly, there have been varying reports of partial or complete transitions to FCO symmetry at low temperatures even in the absence of a magnetic field, leading to phase coexistence.
In single crystals of Mn$_3$O$_4$, Chung et al.\ report an instability along the $[110]$ axis, leading to a partial FCO distortion,\cite{Chung2013b} and Kim et al.\ report a complete transformation to FCO symmetry.\cite{Kim2011a}
Furthermore, in a study of polycrystalline Mn$_3$O$_4$, Kemei et al.\ report a \textgreater50\% FCO phase fraction at low temperatures and propose that a partial FCO transition is intrinsic and may be governed by internal strains.\cite{Kemei2014c}
These varying results raise questions regarding how the ground state properties of Mn$_3$O$_4$ are affected by twin domains, impurities, and other structural properties influenced by specific crystal growth conditions.

Crystals of Mn$_3$O$_4$ are often grown out of a melt ($T_{\text{melt}}>1800$~K); however, such crystals may display twin domains due to the $T=1440$~K cubic-to-tetragonal transition in this material.\cite{Nielsen1969,Yamamoto1972}
The strains near twin domain walls may lead to a partial FCO transition, although this issue has yet to be explored in crystals of Mn$_3$O$_4$.
Furthermore, the misalignment of twin domains prevents an external magnetic field from creating a homogeneous state.
In spectroscopic studies, this inhomogeneity obscures the identification of excitations associated with specific magnetostructural phases.

In this report, we grow untwinned crystals of Mn$_3$O$_4$ from a flux at $T=1370$~K, i.e., below the cubic-to-tetragonal transition.
This allows us to study the characteristic magnons and phonons associated with each magnetostructural phase of Mn$_3$O$_4$ using Raman spectroscopy---a high-resolution ($\Delta E<0.25$~meV) probe of the energies and symmetries of excitations in materials.
We are able to isolate the CDO and FCO phases of Mn$_3$O$_4$ by applying an external magnetic field along the $[010]$ and $[110]$ directions, respectively, and we report the distinct $\mathbf{q}=0$ Raman spectrum of each phase.
We find that the number of modes and the polarization selection rules observed in each phase support the descriptions of the CDO and FCO phases discussed above.
Finally, we obtain Raman spectra of twinned Mn$_3$O$_4$ crystals, grown from a melt for comparison, and we find that twin domains do indeed increase the FCO phase fraction present at low temperatures.

\section{Experiment}

\subsection{Sample preparation}

Following the work of Nielsen,\cite{Nielsen1969} untwinned single crystals of Mn$_3$O$_4$ were grown at the University of Illinois using a flux method.
Commercial powders of manganese(II,III) oxide (Alfa-Aesar, 99.997\%) and anhydrous sodium tetraborate (`Borax', Alfa-Aesar, 99.5\%) were combined in a one-to-one molar ratio, wet milled under ethanol for 12 hours using a FRITSCH Pulverisette P6 planetary ball mill, and dried in air at 150~$^{\circ}$C to remove the ethanol.
To control vaporization loss of Borax at high temperatures, the `crystalline seal' method developed by Wanklyn was used.\cite{Wanklyn1981}
A double crucible was filled with the milled Mn$_3$O$_4$-Borax powder and the top was loosely sealed with a crimped-on lid.
The double crucible was heated in a Lindberg/Blue box furnace to 1100~$^{\circ}$C over 10~hours, held at that temperature for 2~hours, cooled to 900~$^{\circ}$C over 12~hours, and then decanted as a standard `hot-pour'.
The Mn$_3$O$_4$-Borax mixture that sealed the two crucibles together was dissolved by soaking the mixture alternately in deionized water and hydrochloric acid.

The untwinned Mn$_3$O$_4$ crystals were characterized with magnetization and heat capacity measurements (data shown in Appendix~\ref{appendix:characterization}), which show magnetic transitions at $T =$~42, 40, and 34~K, in good agreement with published data.\cite{Jensen1974,Chardon1986}
The crystallographic orientations of the samples used in this experiment were determined via x-ray diffraction measurements performed at room temperature.

The resultant crystals are long, thin parallelepipeds, with approximate dimensions $1\times1\times5$~mm.
The large crystal facets are parallel to $(101)$ and $(011)$ planes, while the long axis is parallel to the $[\bar{1}\bar{1}1]$ direction.
The facets occasionally display visible faults, and x-ray diffraction reveals that there are two twin domains present in these crystals.
The twins are related by a $\pi$ rotation about the long axis, and optical polarization studies confirm that the faults separate the two different crystallographic orientations.
The same type of twinning has previously been observed on a microscopic scale in Mn$_3$O$_4$ grown at temperatures above the cubic-to-tetragonal transition.\cite{Couderc1994}
The twin domains of the flux-grown crystals studied in this report are macroscopic (several~mm in size) and much larger than the laser spot size ($\sim50$~$\mu$m) used in the Raman scattering experiments performed here.
Consequently, we were able to obtain Raman spectra from single-domain regions of these samples; we therefore refer to these crystals as `untwinned' or `single-domain' crystals. 

For comparison, microscopically twinned crystals of Mn$_3$O$_4$ were grown from a melt using a floating-zone method described by Kim et al.\cite{Kim2011a}
The structural and magnetic properties of the resulting crystals are reported elsewhere.\cite{Kim2010, Kim2011a, Kim2011b}
The crystals exhibit three magnetic transitions at $T =$ 43, 39, and 33~K, consistent with previously reported temperatures.\cite{Jensen1974, Chardon1986}
These samples did not exhibit polarization selection rules in Raman scattering measurements, indicating that the twin domains in these samples are smaller than the laser spot size ($\sim50$~$\mu$m) used in the Raman scattering experiments; we therefore refer to these crystals as `twinned' crystals.

\subsection{\label{Geometries}Raman scattering measurements}

Raman scattering measurements were performed using the 647.1~nm excitation line of a krypton ion laser.
The incident laser power was limited to 10~mW and was focused to a $\sim50$-$\mu$m diameter spot to minimize laser heating of the sample, which was estimated to be roughly 4~K.
The scattered light from the samples was collected in a backscattering geometry, dispersed through a triple-stage spectrometer, and detected with a liquid-nitrogen-cooled CCD detector.
The incident light polarization was selected with a polarization rotator, and the scattered light polarization was analyzed with a linear polarizer, providing symmetry information about the excitations studied.
The samples were inserted into a continuous He-flow cryostat, which was horizontally mounted in the open bore of a superconducting magnet, allowing Raman measurements in the temperature range $4\leq T \leq 300$~K and the magnetic field range $0 \leq H \leq 6$~T.

\begin{figure}[h]
	\includegraphics{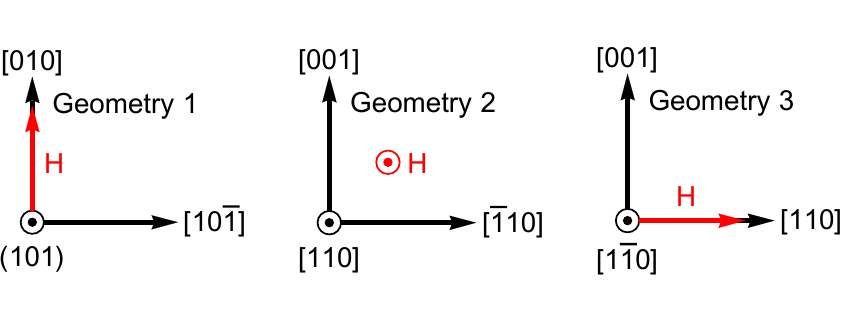}
	\caption{\label{geometries} (Color online) Raman scattering geometries used in variable-magnetic-field measurements.  Each geometry specifies the magnetic field orientation and scattering plane.}
\end{figure}

Three different scattering geometries were employed in the variable-magnetic-field measurements (see Fig.~\ref{geometries}).
Each geometry shown in Fig.~\ref{geometries} specifies the magnetic field orientation and the plane containing the incident- and scattered-light polarizations (scattering plane).
Geometry 1 (Fig.~\ref{geometries}, left) utilizes the as-grown crystal face, which presents a $(101)$ (or, equivalently, $(011)$) surface, and contains a $[010]$ ($[100]$) axis.
A magnetic field was applied in the scattering plane along the $[010]$ ($[100]$) direction. The $[010]$ and $[100]$ directions become magnetically inequivalent in the magnetically-ordered state, with $[010]$ labeling the magnetization direction by convention.\cite{Jensen1974}
Accordingly, when a sufficiently large magnetic field
is applied along the $[010]$ ($[100]$) axis to form a single magnetic domain,\cite{domainalignment} we refer to that direction as the $[010]$ direction.

Crystals were also cut and polished to present a face normal to the $[110]$ (or, equivalently, $[1\bar{1}0]$) direction.
A magnetic field was applied either along the normal of the scattering plane (Geometry 2; see Fig.~\ref{geometries}, middle) or in the scattering plane along the $[\bar{1}10]$ ($[110]$) direction (Geometry 3; see Fig.~\ref{geometries}, right).
It is known that a sufficiently large magnetic field applied along the $[110]$ ($[\bar{1}10]$) axis in the magnetically ordered state causes these directions to become magnetically and structurally inequivalent, as the crystal elongates along the direction of the applied magnetic field.\cite{Nii2013}
We refer to this elongated direction as the $[110]$ direction.

\section{Results and discussion}

\begin{figure}[h]
\includegraphics{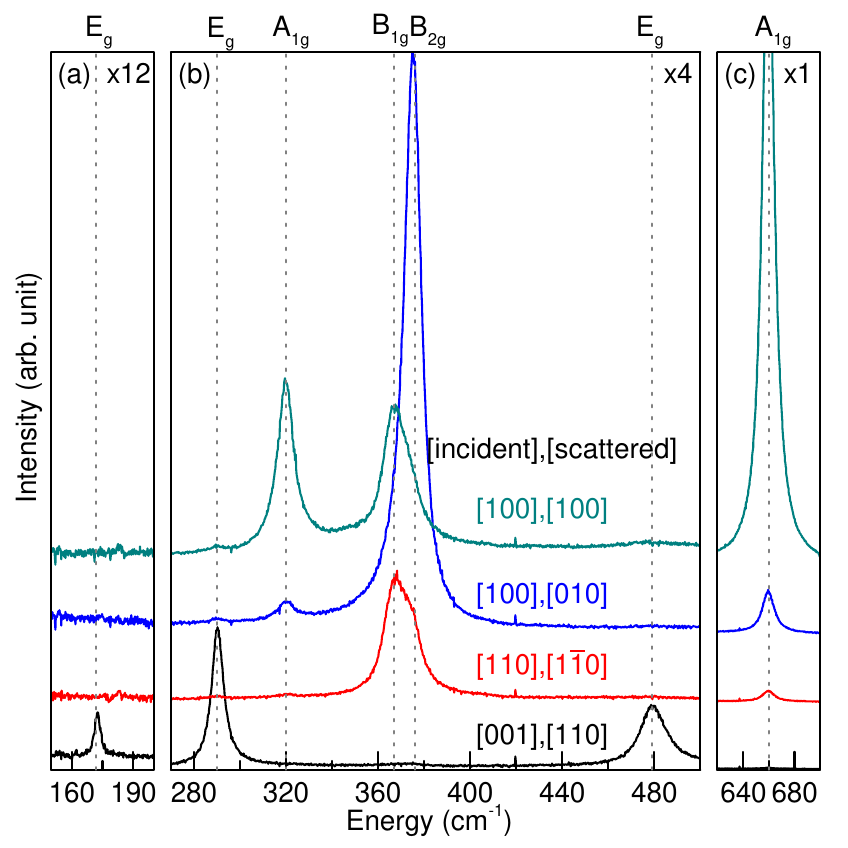}
\caption{\label{tetragonal_polarization} (Color online) Room-temperature Raman scattering spectra of Mn$_3$O$_4$ in the energy ranges (a)~$150-200 \textrm{ cm}^{-1}$, (b)~$270-500 \textrm{ cm}^{-1}$, and (c)~$620-700 \textrm{ cm}^{-1}$ at various polarizations of the incident ([hkl]) and scattered light ([h'k'l']), as indicated by [hkl],[h'k'l']. Phonon symmetries are specified above vertical, dashed lines that indicate the phonon peak positions. The data have been offset for clarity.}
\end{figure}

\subsection{Room-temperature phonon spectra}
Figure~\ref{tetragonal_polarization} presents the $T = 300$~K Raman spectra of an untwinned Mn$_3$O$_4$ crystal taken with various polarizations of the incident and scattered light.
We observe seven of the ten Raman-active phonons predicted by a group theoretical analysis of the Mn$_3$O$_4$ crystal structure using the $D_{4h}$ point group.
The presence of polarization selection rules indicate well-defined crystallographic axes, as expected of an untwinned crystal, and the phonons exhibit symmetries consistent with the $D_{4h}$ point group.
To our knowledge, this is the first experimental determination of the Mn$_3$O$_4$ phonon symmetries.

In addition to demonstrating the absence of twinning in the crystals studied here, the phonon symmetry analysis in Fig.~\ref{tetragonal_polarization} shows for the first time that the large tetragonal distortion from cubic symmetry ($c/a = 1.16 \sqrt{2}$) is manifest in the phonon Raman spectra.
For example, we observe a 54~cm$^{-1}$ (6.75~meV) splitting between the $\omega = 320$~cm$^{-1}$ $A_{1g}$ and $\omega = 374$~cm$^{-1}$ $B_{2g}$ phonons at room temperature; these modes are expected to be degenerate above the cubic-to-tetragonal phase transition at $T = 1440$~K.\cite{*[{See, for example, Table 1 in }] [{}] Takubo2011}

\subsection{\label{Temp_Section}Temperature-dependent excitation spectra}

\begin{figure}[h]
\includegraphics{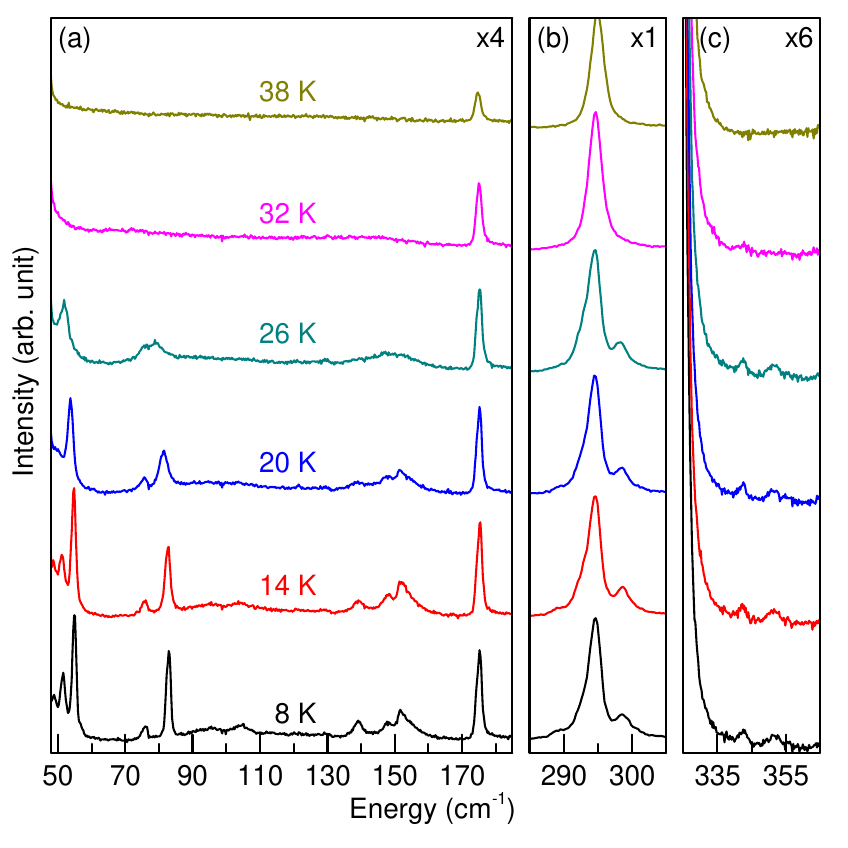}
\caption{\label{zfc} (Color online) Raman-scattering spectra of Mn$_3$O$_4$ at various temperatures in the energy ranges (a)~$50-185 \textrm{ cm}^{-1}$, (b)~$285-305 \textrm{ cm}^{-1}$, and (c)~$325-365 \textrm{ cm}^{-1}$. The data have been offset for clarity.}
\end{figure}

Figure~\ref{zfc} shows the temperature dependence of the untwinned Mn$_3$O$_4$ Raman spectrum (a) in the $50 \leq \omega \leq 185$~cm$^{-1}$ energy range, (b) near the 295~cm$^{-1}$ $E_g$ phonon, and (c) in the $325 \leq \omega \leq 365$~cm$^{-1}$ energy range.
There are three salient developments in Fig.~\ref{zfc} for $T < 32$~K: (1)~Eight new peaks develop in the low-energy range $\omega < 170$~cm$^{-1}$, which we attribute to one-magnon excitations following the analysis of Gleason et al.\cite{Gleason2014}
(2)~Two new phonon peaks appear at $\omega = 342$~cm$^{-1}$ and $\omega = 351$~cm$^{-1}$, whose energies and intensities are consistent with optical phonon modes that have been folded to the Brillouin zone center by a cell-doubling structural modulation.
(3)~The peak at 295~cm$^{-1}$ develops additional structure, indicating a change in the crystal lattice structure discussed further below.

The first two features offer strong confirmation that simultaneous cell-doubling modulations of the magnetic and structural lattices develop below $T = 32$~K.
Previous studies of Mn$_3$O$_4$ observed the appearance of both magnetic\cite{Jensen1974,Nii2013} and nuclear\cite{Nii2013} superlattice reflections at $(h\ k\text{+}\frac{1}{2}\ 0)$ for integer $h,k$ when $T < 34$~K.
Figure~\ref{zfc} shows that, for $T < 32$~K, there is indeed a concomitant development of more than six  magnon peaks and two new phonon peaks---consistent with enlarged magnetic and structural unit cells.
These data provide direct spectroscopic evidence of simultaneous doubling of the magnetic and nuclear unit cells.

\begin{figure}[h]
\includegraphics{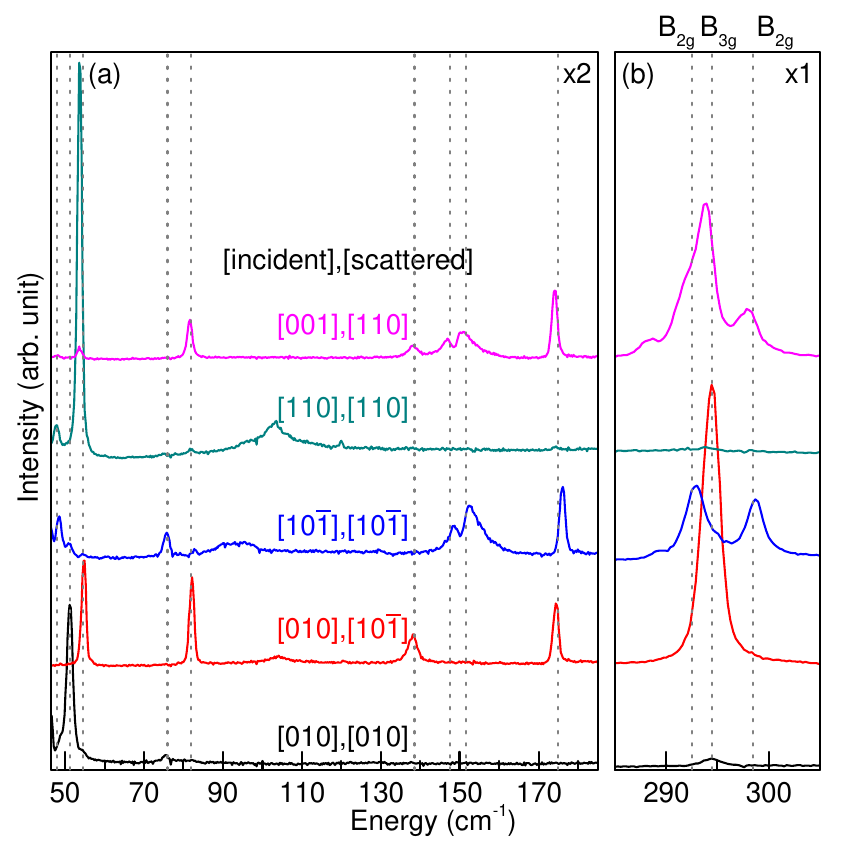}
\caption{\label{doubled_polarization} (Color online) Raman scattering spectra of Mn$_3$O$_4$ in the CDO phase at $T = 8 $~K in the energy ranges (a)~$50-185 \textrm{ cm}^{-1}$ and (b)~$285-305 \textrm{ cm}^{-1}$ taken with various polarizations of the incident ([hkl]) and scattered light ([h'k'l']), as indicated by [hkl],[h'k'l']. Vertical, dashed lines indicate magnon and phonon peak positions in the CDO phase. Phonon symmetries in (b) are specified above their respective lines. The data have been offset for clarity.}
\end{figure}

To clarify the last feature of Fig.~\ref{zfc} concerning a change in the crystal lattice for $T < 32$~K, we performed a polarization analysis of the magnetic and vibrational excitations.
The full details of the symmetry analysis are given in Appendix~\ref{appendix:polarization}, and the experimentally determined energies and symmetries of the observed excitations are tabulated in Table~\ref{table:symmetries}.
Figure~\ref{doubled_polarization} shows representative spectra from this study.
The symmetry selection rules exhibited by the phonon and magnon modes in Fig.~\ref{doubled_polarization} indicate that the low-temperature, zero-field-cooled phase of untwinned Mn$_3$O$_4$ has orthorhombic symmetry with $a$, $b$, and $c$ axes along the $[100]$, $[010]$, and $[001]$ directions, respectively.

The high-symmetry directions we observe are consistent with the modulation of the Mn$^{3+}_4$ tetrahedra proposed by Nii et al.\ and illustrated in Fig.~\ref{lattice}(b).\cite{Nii2013}
We note that the displacement pattern shown in Fig.~\ref{lattice}(b) removes the twofold rotations about the $[110]$ and $[\bar{1}10]$ axes, as well as any mirror operations in planes orthogonal to these directions.
On the other hand, the twofold rotations about the $[100]$ and $[010]$ axes are preserved (up to a fractional translation, in the case of the $[010]$ axis).

We clearly resolve three non-degenerate phonons near 295~cm$^{-1}$ in the CDO phase (see Fig. \ref{doubled_polarization}(b)), including a $B_{3g}$ mode ($\omega = 295$~cm$^{-1}$) and two $B_{2g}$ modes ($\omega = 293$ and 298~cm$^{-1}$).
The tetragonal-to-orthorhombic transition is expected to split the $\omega = 295$~cm$^{-1}$ $E_g$ phonon into a $B_{3g}$ phonon and a \emph{single} $B_{2g}$ phonon.
The presence of a second $B_{2g}$ phonon is associated with a phonon mode that has been folded to $\mathbf{q}=0$ because of the structural modulation for $T < 32$~K, similar to the folded phonon modes at $\omega =$ 342 and 351 cm$^{-1}$ discussed previously.

Based on the number of modes we observe and their respective polarization selection rules, we conclude that untwinned Mn$_3$O$_4$ transitions to the CDO phase depicted in Fig.~\ref{lattice}(b) for $T < 32$~K in the absence of a magnetic field.
An explanation for reports of partial or complete transitions to an FCO phase at low temperatures is given in \S{}\ref{Twinning}.

\subsection{Magnetic-field-dependent excitation spectra}

Previous studies\cite{Kim2010,Kim2011a,Nii2013} have shown that an applied magnetic field can significantly influence the crystal lattice of Mn$_3$O$_4$, and that the structural response depends on the orientation of the applied field.
For example, a recent x-ray and neutron scattering study by Nii et al.\ reported that a magnetic field applied along the $[110]$ direction stabilizes the FCO phase with $a$, $b$, and $c$ axes along the $[1\bar{1}0]$, $[110]$, and $[001]$ directions, respectively.\cite{Nii2013}
Additionally, no satellite peaks were observed in neutron and x-ray diffraction measurements of the FCO phase, indicating the structural and magnetic modulations are absent in this phase.\cite{Nii2013}
Our spectroscopic studies of untwinned Mn$_3$O$_4$ allow a further elucidation of the magnetic-field-induced phases of Mn$_3$O$_4$, by allowing us to study \textit{simultaneously} the magnetic and phonon excitation spectra while applying a magnetic field along well-defined crystallographic directions.
We will first show that a magnetic field applied along the $[110]$ direction does indeed lead to a transition to the FCO magnetostructural phase.
Next, we will show that a magnetic field applied along the $[010]$ direction preserves the CDO magnetostructural phase.
However, we will also show that reversing the magnetic field from $[010]$ to $[0\bar{1}0]$ induces a CDO$\rightarrow$FCO$\rightarrow$CDO transition.

\subsubsection{\label{H parallel [110]}$\mathbf{H} \parallel [110]$: face-centered-orthorhombic phase}

\begin{figure}[h!]
\includegraphics{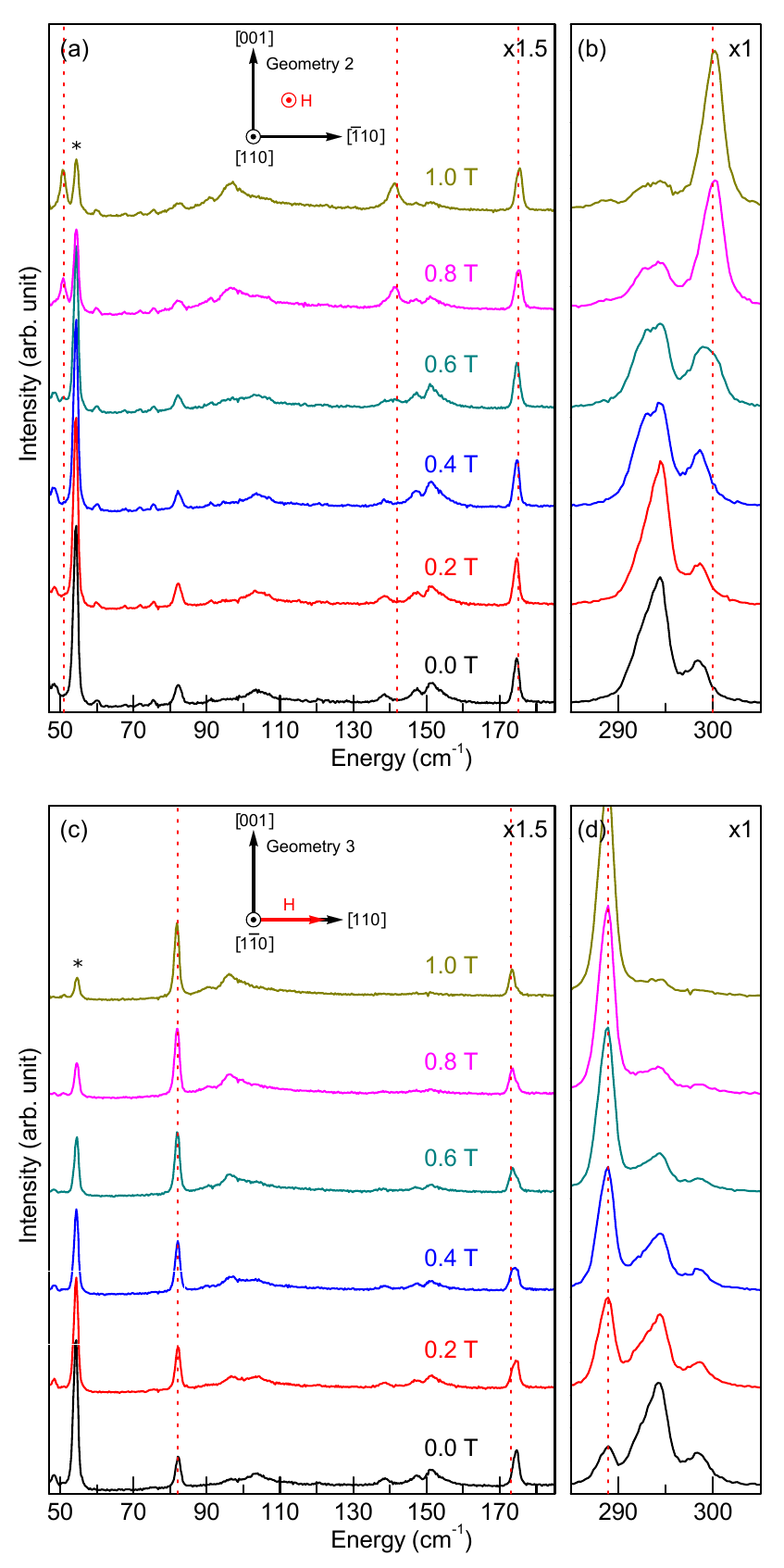}
\caption{\label{undoubled_field} (Color online) Raman scattering spectra of Mn$_3$O$_4$ at $T = 8$~K taken in (a),(b)~Geometry 2 and (c),(d)~Geometry 3 
(see \S{}\ref{Geometries}) at various magnetic 
fields in the energy ranges 
(a),(c) $50-185 \textrm{ cm}^{-1}$ and 
(b),(d) $285-305 \textrm{ cm}^{-1}$. Vertical, dashed lines indicate magnon and phonon peak positions in the FCO phase. The asterisk indicates a peak associated with a small, remnant CDO phase fraction.\cite{CDOremnant} The data have been offset for clarity.}
\end{figure}

Figure~\ref{undoubled_field} shows the magnetic-field dependence of the Raman spectrum in the energy ranges $50 \leq \omega \leq 185$~cm$^{-1}$ and $285 \leq \omega \leq 305$~cm$^{-1}$ for a magnetic field applied along the $[110]$ direction measured in (a),(b) Geometry 2 and (c),(d) Geometry 3 at $T = 8$~K.
The two scattering geometries are employed to allow the observation of nearly all of the Raman-active modes.
For all measurements, the sample was zero-field-cooled from room temperature to $T = 8$~K before a magnetic field was applied.
The resulting $H = 0$~T and $T = 8$~K Raman spectra (bottom curves in Figs.~\ref{undoubled_field}(a) and \ref{undoubled_field}(b)) are consistent with the CDO phase, as discussed in \S{}\ref{Temp_Section}.\cite{FCOremnant}
When $\mathbf{H} \parallel [110]$, as the magnetic field magnitude is increased, the spectra in Fig.~\ref{undoubled_field} change dramatically in the range $0 \leq H \leq 1$~T, indicating a magnetostructural phase transition.
No further changes in the spectra are observed for $1 < H \leq 6$~T (data not shown).
We show below that the phase stabilized for $H > 1$~T with $\mathbf{H} \parallel [110]$ is consistent with the FCO phase, in agreement with the results of Nii et al.\cite{Nii2013}

\begin{figure}[h]
\includegraphics{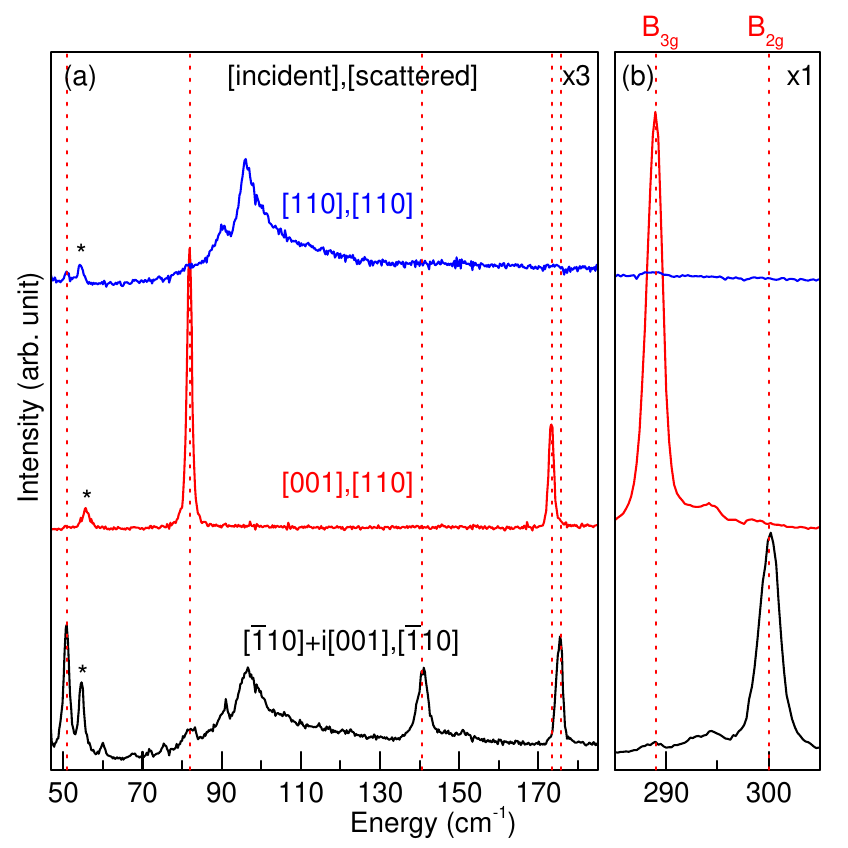}
\caption{\label{undoubled_polarization} (Color online) Raman scattering spectra of Mn$_3$O$_4$ in the FCO phase at $T = 8 $~K and $H = 1$~T along the $[110]$ direction in the energy ranges (a)~$50-185 \textrm{ cm}^{-1}$ and (b)~$285-305 \textrm{ cm}^{-1}$ taken with various polarizations of the incident ([hkl]) and scattered light ([h'k'l']), as indicated by [hkl],[h'k'l']. ($[h_1k_1l_1] + i[h_2k_2l_2]$ denotes circular polarization.) Vertical, dashed lines indicate magnon and phonon peak positions in the FCO phase. Phonon symmetries in (b) are specified above their respective lines. The asterisk indicates a peak associated with a small, remnant CDO phase fraction.\cite{CDOremnant}  The data have been offset for clarity.}
\end{figure}

To verify the FCO symmetry of the magnetostructural phase for $H > 1$~T and $\mathbf{H} \parallel [110]$, we performed a polarization study of the magnetic and vibrational excitations in this phase regime.
The full details of the symmetry analysis are given in Appendix~\ref{appendix:polarization}, and the experimentally determined energies and symmetries of the observed excitations are tabulated in Table~\ref{table:symmetries}. Fig.~\ref{undoubled_polarization} shows representative spectra from this study.
The broad, asymmetric peak at $\omega = 96 \text{ cm}^{-1}$ in the FCO phase is identified as a two-magnon excitation based on the analysis of Gleason et al.\ and will not be discussed.\cite{Gleason2014}
The phonon and magnon modes exhibit symmetry selection rules that are consistent with an FCO structure having $a$, $b$, and $c$ axes along the $[1\bar{1}0]$, $[110]$, and $[001]$ directions, respectively. 
These new high-symmetry axes can be understood by examining the displacement pattern proposed by Nii et al.\cite{Nii2013} and depicted in Fig.~\ref{lattice}(c).
The uniform distortion of the Mn$^{3+}_4$ tetrahedra removes the twofold rotational symmetry about the $[100]$ and $[010]$ axes, as well as the mirror operations in planes orthogonal to these directions.
On the other hand, the lattice depicted in Fig.~\ref{lattice}(c) possesses twofold rotational symmetry about the $[110]$ and $[\bar{1}10]$ directions, which is consistent with the selection rules we observe.
Furthermore, the splitting between the two normal modes near 295~cm$^{-1}$ is $\sim11$~cm$^{-1}$ in the FCO phase, which is $\sim4$ times larger than the splitting between the equivalent modes in the CDO phase.
As this splitting reflects the loss of fourfold symmetry in the $(001)$ plane, a larger splitting in the FCO phase makes sense, given the more pronounced asymmetry between the in-plane, high-symmetry directions caused by the uniform distortion of the crystal.

\begin{figure}[h]
\includegraphics{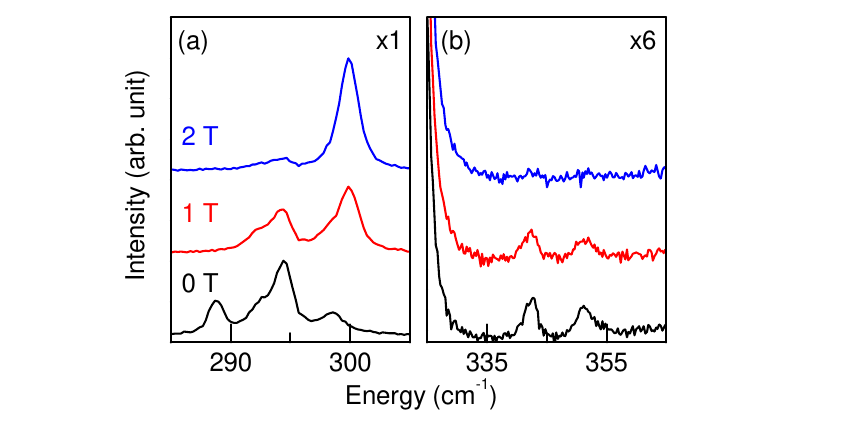}
\caption{\label{undoubled_field_2} (Color online) Raman scattering spectra of Mn$_3$O$_4$ at $T = 8$~K taken in Geometry 2 (see \S{}\ref{Geometries}) at various magnetic fields in the energy ranges 
(a)~$285-305 \textrm{ cm}^{-1}$ and 
(b)~$325-365 \textrm{ cm}^{-1}$. The data have been offset for clarity.}
\end{figure}

To complete the characterization of the FCO phase, we examined the folded optical phonons, which are a signature of the structural cell-doubling modulation.
Figure~\ref{undoubled_field_2} shows the magnetic-field dependence of (a) the phonon spectrum in the $285 \leq \omega \leq 305$~cm$^{-1}$ energy range and (b) the folded phonons near 342~cm$^{-1}$ and 351~cm$^{-1}$ for $\mathbf{H} \parallel [110]$ and $T = 8$~K.
As discussed above, the $H = 0$~T phonon and magnon spectra of untwinned Mn$_3$O$_4$ below $T=32$~K are indicative of the CDO phase.\cite{FCOremnant}
When $\mathbf{H} \parallel [110]$, the intensities of the folded phonons diminish with increasing magnetic field.
For $H > 2$~T, the folded phonon modes vanish as the FCO phase is stabilized, as indicated by the appearance of the $\omega = 300$~cm$^{-1}$ phonon and the suppression of the $\omega =$~293, 295, and 298~cm$^{-1}$ phonons in Fig.~\ref{undoubled_field_2}(a) (compare with field dependence in Fig.~\ref{undoubled_field}(b)).
This field dependence confirms that a magnetic field applied along the $[110]$ direction suppresses the cell-doubling \emph{structural} modulation in Mn$_3$O$_4$, consistent with the results of Nii et al.\cite{Nii2013}
Additionally, we observe fewer than six magnons in the FCO phase, as seen in Fig.~\ref{undoubled_field}, consistent with the number of spins in the FCO unit cell, indicating the \emph{magnetic} cell-doubling modulation is not present in the FCO phase.
These data provide further evidence that the structural and magnetic cell-doubling modulations are absent in the FCO phase.

A summary of the CDO and FCO excitation spectra is given in Table~\ref{table:symmetries}, which tabulates the experimentally determined energies and symmetries of the Raman-active one-magnon and one-phonon excitations with $\omega \leq 300$~cm$^{-1}$ for the CDO and FCO phases. Each normal mode can be assigned to an irreducible representation of the $D_{2h}$ point group.  The details of the analysis performed are given in Appendix~\ref{appendix:polarization}.

\begin{table}[h]
	
	\begin{tabular}{|c|c|c|c|}
		\hline
		\multirow{2}{*}{}   &   \multicolumn{2}{| c |}{Energy (cm$^{-1}$)}   & \multirow{2}{*}{$\Gamma$($D_{2h}$)}   \\ \cline{2-3}
		& CDO phase   &  FCO phase   &             \\ \cline{1-4}
		\multicolumn{1}{ |c | }{\multirow{12}{*}{Magnon}}
		&  49   &   ---  &   A$_{g}$   \\ \cline{2-4}
		&  51   &   51   &   A$_{g}$   \\ \cline{2-4}
		&  54   &   ---  &   B$_{1g}$  \\ \cline{2-4}
		&  ---  &   56   &   B$_{1g}$  \\ \cline{2-4}
		&  75   &   ---  &   A$_{g}$   \\ \cline{2-4}
		&  83   &   83   &   B$_{3g}$  \\ \cline{2-4}
		&  139  &   ---  &   B$_{3g}$  \\ \cline{2-4}
		&  ---  &   142  &   B$_{2g}$  \\ \cline{2-4}
		&  148  &   ---  &   B$_{2g}$  \\ \cline{2-4}
		&  152  &   ---  &   B$_{2g}$  \\ \hline
		\multicolumn{1}{|c|}{\multirow{11}{*}{Phonon}}  
		& ---  &   174  &   B$_{3g}$  \\ \cline{2-4}
		&  175  &   ---  &   B$_{2g}$  \\ \cline{2-4}
		&  175  &   ---  &   B$_{3g}$  \\ \cline{2-4}
		&  ---  &   176  &   B$_{2g}$  \\ \cline{2-4}
		&  ---  &   289  &   B$_{3g}$  \\ \cline{2-4}
		&  293  &   ---  &   B$_{2g}$  \\ \cline{2-4}
		&  295  &   ---  &   B$_{3g}$  \\ \cline{2-4}
		&  298  &   ---  &   B$_{2g}$  \\ \cline{2-4}
		&  ---  &   300  &   B$_{2g}$  \\ \hline
	\end{tabular}

	\caption{\label{table:symmetries}Measured energies and symmetries of the magnetic and vibrational excitations in the CDO and FCO phases of Mn$_3$O$_4$ at $T = 8$~K. See Appendix~\ref{appendix:polarization} for details. The energy difference between the $B_{2g}$ and $B_{3g}$ phonons at $\omega = 175$~cm$^{-1}$ could not be resolved in the CDO phase.}
	
\end{table}

\subsubsection{\label{H parallel [010]}$\mathbf{H} \parallel [010]$: cell-doubled-orthorhombic phase}

\begin{figure}[h]
\includegraphics{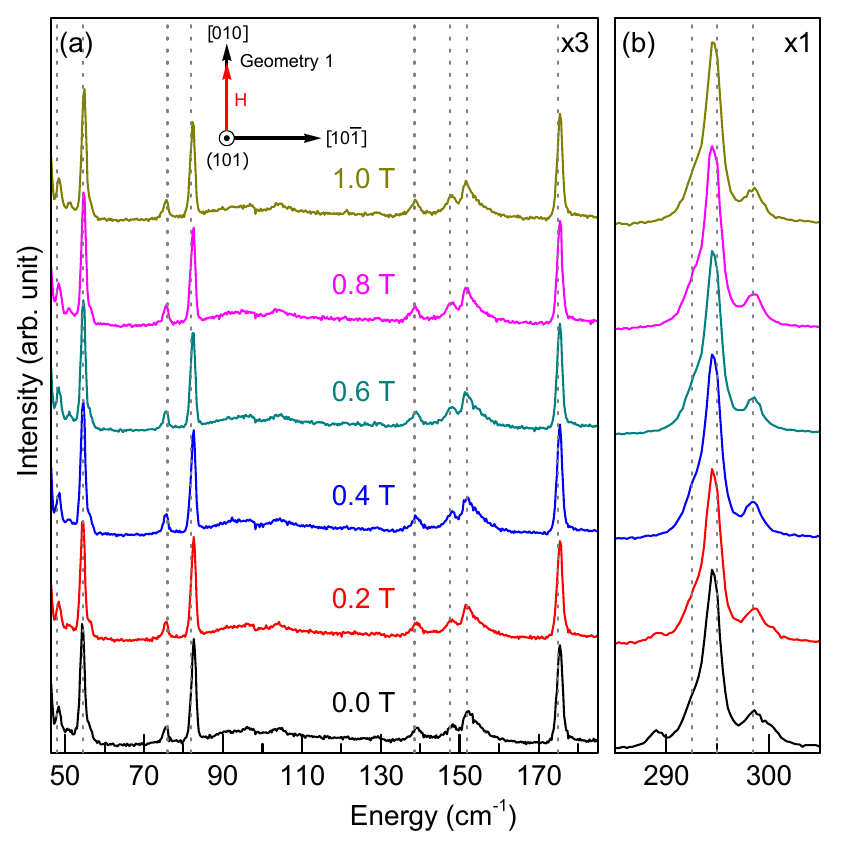}
\caption{\label{doubled_field} (Color online) 
Raman scattering spectra of Mn$_3$O$_4$ at $T = 8$~K taken in Geometry 1 (see \S{}\ref{Geometries}) at various magnetic 
fields in the energy ranges 
(a)~$50-185 \textrm{ cm}^{-1}$ and 
(b)~$285-305 \textrm{ cm}^{-1}$. Vertical, dashed lines indicate magnon and phonon peak positions in the CDO phase. 
The data have been offset for clarity.}
\end{figure}

We next show that a magnetic field applied along the magnetic easy axis $[010]$ preserves the CDO phase in untwinned Mn$_3$O$_4$.
Figure~\ref{doubled_field} shows the magnetic-field dependence of the Raman spectrum (a) in the $50 \leq \omega \leq 185$~cm$^{-1}$ energy range and (b) near 295~cm$^{-1}$ for $\mathbf{H} \parallel [010]$ at $T = 8$~K.
The sample was zero-field-cooled from room temperature to $T = 8$~K before the magnetic field was applied.
Once again, the $H = 0$~T Raman spectrum of untwinned Mn$_3$O$_4$ (bottom curve) is characteristic of the CDO phase.\cite{FCOremnant}

As the magnetic field is increased to $H = 1$~T, the spectrum does not change appreciably.
Moreover, no significant changes are observed in the spectrum for magnetic fields up to $H = 6$~T (data not shown).
We conclude that the CDO phase is stable for a magnetic field applied along the $[010]$ direction in Mn$_3$O$_4$.

Previous Raman scattering studies by Kim et al.\ reported that an FCO phase can be stabilized for magnetic fields applied in the $(001)$ plane.\cite{Kim2010,Kim2011a}  
Indeed, our phonon Raman data are qualitatively similar to those of Kim et al.\cite{Kimfootnote}
However, Kim et al.\ reported that the FCO phase is stabilized for an $H > 3$~T magnetic field oriented either along or orthogonal to the easy axis direction ($[010]$),\cite{Kim2010,Kim2011a} which conflicts with the results in \S{}\ref{H parallel [110]} and \S{}\ref{H parallel [010]} of this paper.  
Our results, as well as the results of Nii et al.,\cite{Nii2013} suggest that magnetic fields that stabilized the FCO phase in refs.~[\onlinecite{Kim2010}] and [\onlinecite{Kim2011a}] were actually along the $[\bar{1}10]$ or $[110]$ directions.  
This discrepancy likely results from an inconsistent use of both face-centered-tetragonal and body-centered-tetragonal settings when describing crystallographic orientations in refs.~[\onlinecite{Kim2010}] and [\onlinecite{Kim2011a}].

\subsubsection{$\mathbf{H} \parallel [010]$: phase hysteresis}

\begin{figure}[h]
\includegraphics{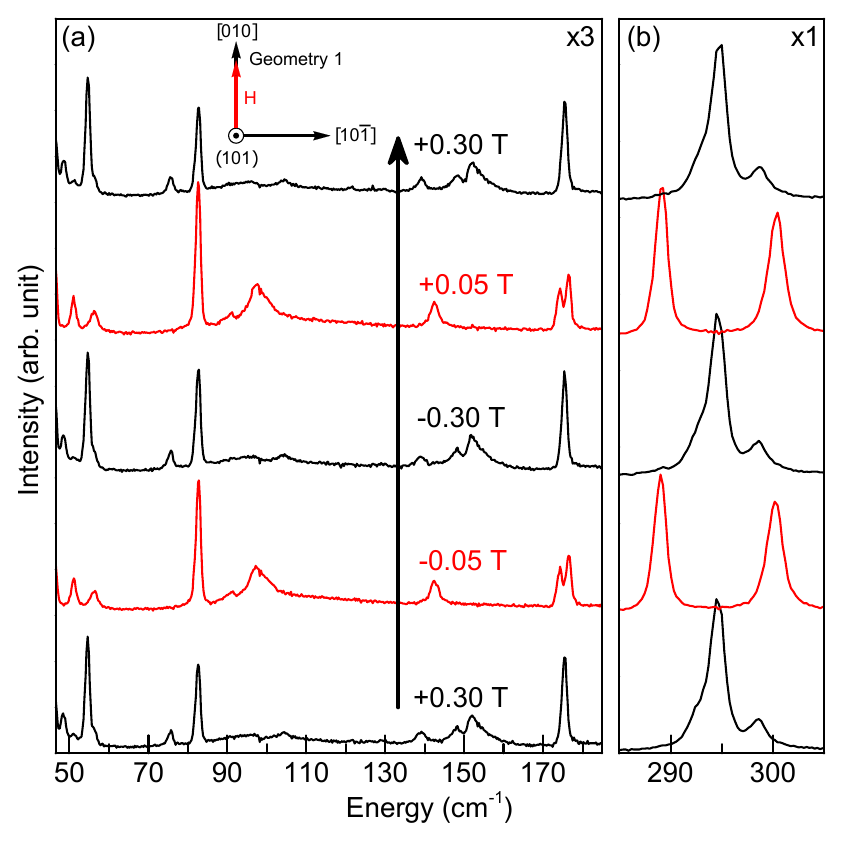}
\caption{\label{hysteresis} (Color online)
Raman scattering spectra of Mn$_3$O$_4$ at $T = 8$~K taken in Geometry 1 (see discussion in \S{}\ref{Geometries}) at various magnetic 
fields in the energy ranges 
(a)~$50-185 \textrm{ cm}^{-1}$ and 
(b)~$285-305 \textrm{ cm}^{-1}$. The Raman spectrum for the CDO and FCO phases are marked in black and red, respectively. The arrow indicates the sequence of field values used. The data have been offset for clarity.}
\end{figure}

After applying a magnetic field along the $[010]$ direction and subsequently reducing the field to $H = 0$~T, we occasionally found that the CDO spectrum would transform to the FCO spectrum.
This motivated us to perform a sequence of measurements in which a magnetic field applied along the $[010]$ direction was cycled between $H = +0.3$ and $H = -0.3$~T.
Figure~\ref{hysteresis} shows the Raman spectra of Mn$_3$O$_4$ for various fields in this sequence.

Figure~\ref{hysteresis} demonstrates that reversing a magnetic field along the $[010]$ direction can dramatically influence the phase fraction of Mn$_3$O$_4$.
We have already identified and attributed the $H = \pm 0.3$~T Raman spectra (marked in black) to the CDO phase.
The red spectrum observed at $H = \pm 0.05$~T is exactly the FCO spectrum formerly obtained by applying a $H = 1$~T magnetic field along the $[110]$ direction, which can be seen by comparing the peaks in this spectrum with those in the $H = 1$~T spectra in Fig.~\ref{undoubled_field}.

Note that we previously demonstrated that a magnetic field along the $[010]$ direction stabilizes the CDO phase.
Remarkably, if the magnetic field is applied along the $[010]$ direction and subsequently reversed to the $[0\bar{1}0]$ direction, the phase fractions change dramatically near $H = 0$~T.
In fact, the crystal intermediately transitions fully to the FCO phase before reverting to the CDO phase at higher field strengths.
The ability to tune between the CDO and FCO phases with very small magnetic fields along the $[010]$ direction further illustrates that the free energies of these two phases are nearly degenerate in Mn$_3$O$_4$ at $T = 8$~K.

\subsection{\label{Twinning}Effect of twin domains}

We have identified the distinct Raman spectra of the CDO and FCO phases of Mn$_3$O$_4$ at low temperatures.
The near degeneracy of these phases is evidenced by the ability to manipulate the phase fraction at low temperature with modest magnetic fields applied in the $(001)$ plane.
Consequently, it is likely that other parameters, e.g., strain, can also influence the phase fraction of Mn$_3$O$_4$.
Indeed, an x-ray diffraction study by Kemei et al.\ reported phase coexistence in polycrystalline Mn$_3$O$_4$ at low temperatures, which was attributed to strain arising from domain walls.\cite{Kemei2014c}

It is well established that growth temperature contributes to the presence or absence of structural domains in crystals of Mn$_3$O$_4$.
As discussed in \S{}\ref{introduction}, twin domains can be introduced in  Mn$_3$O$_4$ crystals by growing at temperatures above the $T = 1440$~K cubic-to-tetragonal Jahn-Teller transition, e.g., by growing crystals from a melt.\cite{Nielsen1969,Yamamoto1972}
On the other hand, untwinned Mn$_3$O$_4$ crystals can be obtained by growing at temperatures below the Jahn-Teller transition, e.g., from a flux\cite{Nielsen1969} or via chemical vapor transport\cite{Yamamoto1972}.

\begin{figure}[h]
\includegraphics{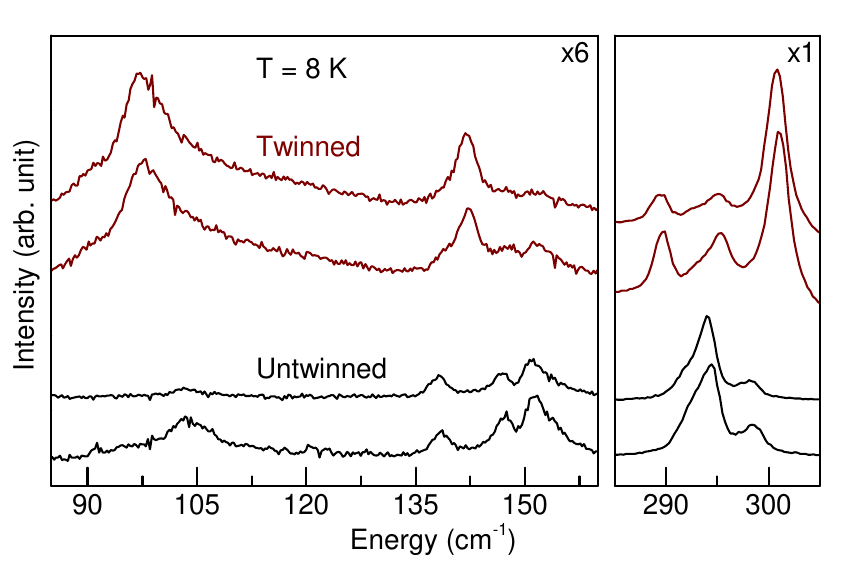}
\caption{\label{phase_fraction} (Color online)
Raman scattering spectra of untwinned and twinned Mn$_3$O$_4$ crystals at $T = 8$~K in the energy ranges 
(a)~$85-160 \textrm{ cm}^{-1}$ and 
(b)~$285-305 \textrm{ cm}^{-1}$. The data have been offset for clarity.}
\end{figure}

To study the effect of structural domains on the low-temperature phase fraction in Mn$_3$O$_4$ crystals, we compared Raman scattering spectra from both untwinned and twinned Mn$_3$O$_4$ crystals in the absence of a magnetic field.
Figure~\ref{phase_fraction} shows the zero-field-cooled, $T = 8$~K Raman spectra of multiple untwinned and twinned Mn$_3$O$_4$ crystals.
The untwinned crystals overwhelmingly exhibit the CDO phase, while the twinned crystals exhibit a coexistence of both CDO and FCO phases.
This indicates that the low-temperature FCO phase fraction is strongly influenced by the presence of structural domains in Mn$_3$O$_4$ crystals, with structural domains promoting the nucleation of the FCO phase.

As the anisotropic responses of the electronic and structural properties of Mn$_3$O$_4$ to an applied magnetic field are thought to result from tuning between the CDO and FCO phases, controlling the phase fraction via the introduction of structural domains may present a novel mechanism for controlling the functional behavior of Mn$_3$O$_4$.

\section{Summary}

We performed a variable-temperature and variable-magnetic-field Raman scattering study of untwinned Mn$_3$O$_4$ crystals.  
By studying untwinned crystals, we are able to identify the distinctive $\mathbf{q}=0$ magnetic and vibrational Raman excitation energies and symmetries of Mn$_3$O$_4$ for both the CDO and FCO phases at $T = 8$~K.  
Our spectroscopic results for each phase support the microscopic description of the two magnetostructural phases given by Nii et al.\cite{Nii2013} 
We also report the effects of a magnetic field applied in the $(001)$ plane on the magnetostructural phase mixture of Mn$_3$O$_4$ at $T = 8$~K.
Specifically, we find that field orientations $\mathbf{H} \parallel [010]$ and $\mathbf{H} \parallel [110]$ for $H>1$~T select the CDO and FCO phases, respectively, in untwinned Mn$_3$O$_4$. 
Lastly, we compare the $H=0$~T, $T = 8$~K Raman spectra of untwinned crystals to the spectra of twinned crystals of Mn$_3$O$_4$.  
We find that the presence of twin domains increases the low-temperature phase fraction of the FCO phase in crystals of Mn$_3$O$_4$.  
This result may explain the varying reports of partial or complete transitions to the FCO phase and also suggests a mechanism for controlling the low-temperature phase and functional behavior of Mn$_3$O$_4$.

\begin{acknowledgments}

Research by T.B., S.L.G., and S.L.C.\ was supported by the National Science Foundation under Grant NSF DMR 14-64090. 
Research by A.T. and G.J.M.\ was supported by the NSF under Grant DMR-1455264-CAR.

\end{acknowledgments}

\appendix

\section{\label{appendix:characterization}Sample characterization}

\begin{figure}[t]
	\centering
	\includegraphics[width = 0.9\linewidth]{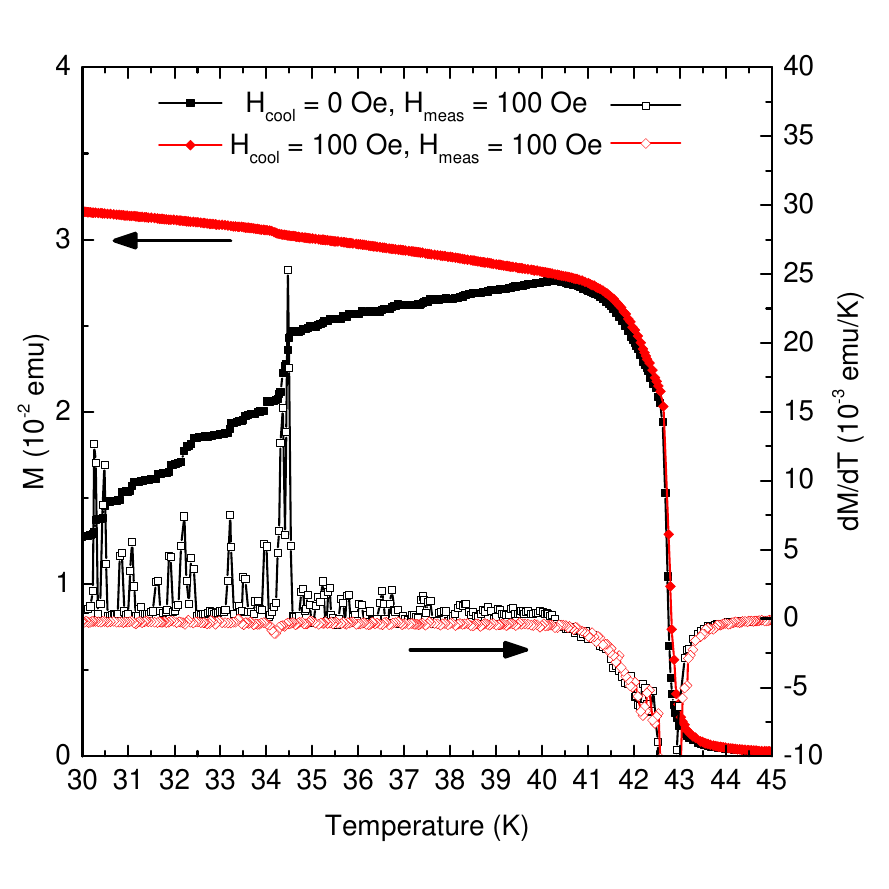}
	\caption{(Color online) DC magnetization measurements, including both zero-field-cooled (black) and field-cooled (red) data. Closed and open symbols correspond to magnetization and its temperature derivative, respectively.}
	\label{fig:magnetization}
\end{figure}

\begin{figure}[h]
	\centering
	\includegraphics[width = 0.9\linewidth]{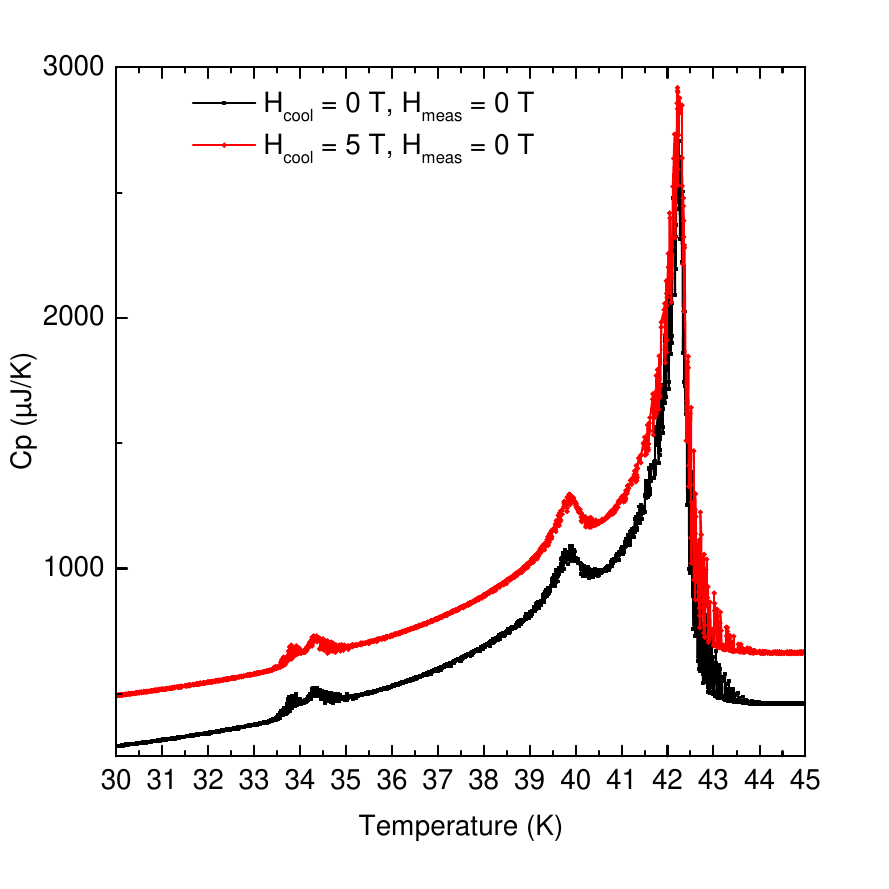}
	\caption{(Color online) Heat capacity measurements, including both zero-field-cooled (black) and field-cooled (red) data.}
	\label{fig:heatcapacity}
\end{figure}

We collected DC magnetization data using a Quantum Design MPMS-3 (see Fig.~\ref{fig:magnetization}), as well as heat capacity data using a Quantum Design PPMS-DynaCool (see Fig.~\ref{fig:heatcapacity}). Both zero-field-cooled and field-cooled data are shown.
The data show magnetic transitions at $T =$~42, 40, and 34~K, in good agreement with published data.\cite{Jensen1974,Chardon1986}
The apparent 0.5~K splitting of the $T = 34$~K cell-doubling transition in Fig.~\ref{fig:heatcapacity} is an artifact of the method by which we collected our data---one peak is from the warming part of the cycle, while the other is from the cooling section. 
As discussed by Guillou et al., this is a clear signature of a first-order phase transition.\cite{Guillou2011} 
It also suggests that the quality of these crystals is high, as low-quality, multi-grain crystals have transition temperatures which are too smeared out to see such sharp features.

\section{\label{appendix:polarization}Polarization studies}

\begin{figure}[h!]
	\includegraphics[width = 1.0\linewidth]{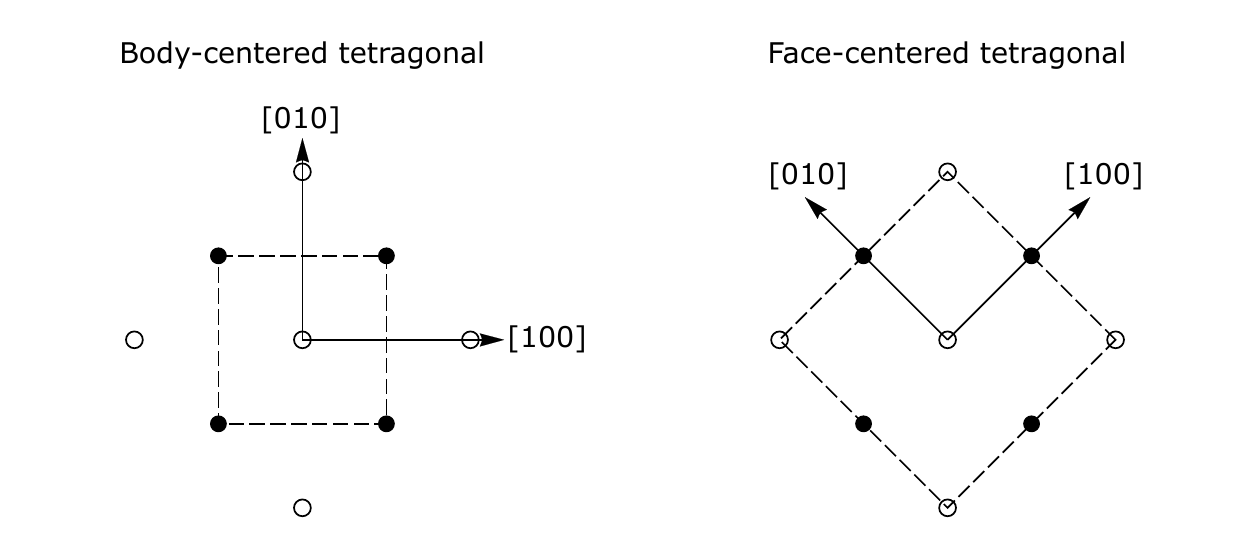}
	\caption{\label{fig:settings}Illustration showing the equivalence of body-centered-tetragonal and face-centered-tetragonal Bravais lattices. White and black circles denote lattice points at different \textit{c}-axis positions.}
\end{figure}

At room temperature, the Bravais lattice of Mn$_3$O$_4$ is centered tetragonal. 
Two equivalent conventional unit cells are employed in the literature: a body-centered-tetragonal cell (Fig.~\ref{fig:settings}, left) and a face-centered-tetragonal cell (Fig.~\ref{fig:settings}, right).
These coordinate systems differ by a $\pi/4$ rotation about the $[001]$ axis.
In this paper, we choose the body-centered-tetragonal setting as our coordinate system.

\begin{figure}[h!]
	\includegraphics[width = 1.0\linewidth]{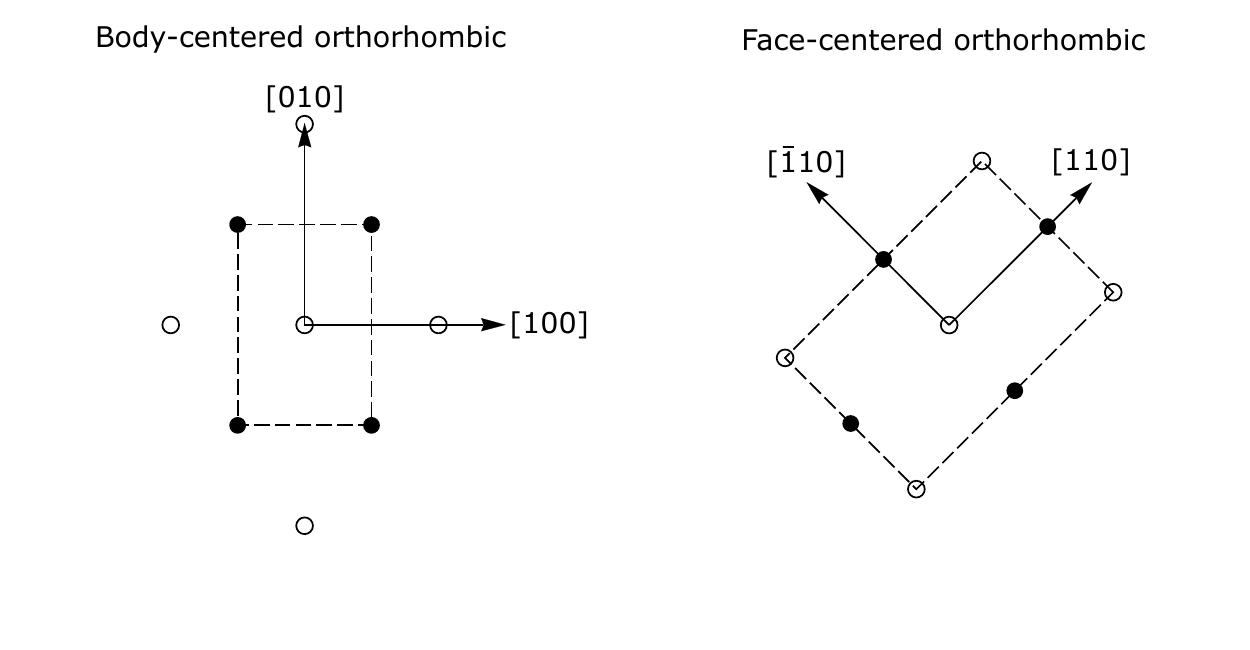}
	\caption{\label{fig:FCO_BCO_distortion}Illustration of body-centered-orthorhombic and FCO Bravais lattices. White and black circles denote lattice points at different \textit{c}-axis positions. Crystallographic directions are with respect to the body-centered-tetragonal coordinate system.}
\end{figure}

At room temperature, the Mn$_3$O$_4$ crystal structure possesses 4-fold rotational symmetry about the principal axis $[001]$, up to a fractional translation.
The crystal structure also has two pairs of 2-fold symmetry axes in the $(001)$ plane: the secondary axes $[100]$ and $[010]$ and the tertiary axes $[\bar{1}10]$ and $[110]$.

The low-temperature, orthorhombic magnetostructural phases of Mn$_3$O$_4$, which form the subject of this paper, are distorted such that 4-fold symmetry is lost.
Each phase retains only \emph{one} pair of 2-fold symmetry axes in the $(001)$ plane.
The cell-doubled-orthorhombic (CDO) phase retains 2-fold symmetry about the $[100]$ and $[010]$ directions (see Fig.~\ref{fig:FCO_BCO_distortion}, left).
(Because the cell-doubling modulation present in this phase leads to a unit cell which is not internally centered, we refer to this phase as cell-doubled orthorhombic rather than body-centered orthorhombic.)
On the other hand, the face-centered-orthorhombic (FCO) phase retains 2-fold symmetry about the $[\bar{1}10]$ and $[110]$ directions (see Fig.~\ref{fig:FCO_BCO_distortion}, right).

Table~\ref{table:tensors} presents the $D_{2h}$ Raman tensors for each phase with respect to the body-centered-tetragonal coordinate system.
Symmetric tensors were found sufficient to describe the polarization dependences observed for each peak's intensity.
Magnetic excitations are assigned to irreducible representations of the $D_{2h}$ point group, rather than the monoclinic magnetic point group reported by Jensen et al.,~\cite{Jensen1974} for the same reason.

\begin{table*}[t]

\begin{tabular}{|cccc|cccc|}
	\hline
	\multicolumn{4}{|c}{CDO Raman tensors} & \multicolumn{4}{|c|}{FCO Raman tensors}  \\
	\hline
	$A_g$  &  $B_{1g}$  &  $B_{2g}$  &  $B_{3g}$  &   $A_g$  &  $B_{1g}$  &  $B_{2g}$  &  $B_{3g}$  \\ 
	$\left( \begin{array}{ccc}  a & 0 & 0 \\ 0 & b & 0 \\ 0 & 0 & c   \end{array} \right)$   &  
	$\left( \begin{array}{ccc}  0 & d & 0 \\ d & 0 & 0 \\ 0 & 0 & 0   \end{array} \right)$   &   
	$\left( \begin{array}{ccc}  0 & 0 & e \\ 0 & 0 & 0 \\ e & 0 & 0	\end{array} \right)$   &  
	$\left( \begin{array}{ccc} 0 & 0 & 0 \\ 0 & 0 & f \\ 0 & f & 0	\end{array} \right)$   &
	$\left( \begin{array}{ccc}  \frac{a+b}{2} & \frac{b-a}{2} & 0 \\ \frac{b-a}{2} & \frac{a+b}{2} & 0 \\ 0 & 0 & c   \end{array} \right)$   &
	$\left( \begin{array}{ccc}  d & 0 & 0 \\ 0 & -d & 0 \\ 0 & 0 & 0   \end{array} \right)$   &
	$\left( \begin{array}{ccc}  0 & 0 & \frac{e}{\sqrt{2}} \\ 0 & 0 & \frac{e}{\sqrt{2}} \\ \frac{e}{\sqrt{2}} & \frac{-e}{\sqrt{2}} & 0   \end{array} \right)$   &
	$\left( \begin{array}{ccc}  0 & 0 & \frac{f}{\sqrt{2}} \\ 0 & 0 & \frac{f}{\sqrt{2}} \\ \frac{f}{\sqrt{2}} & \frac{f}{\sqrt{2}} & 0  \end{array} \right)$ \\ \hline 

\end{tabular}

	\caption{\label{table:tensors}$D_{2h}$ Raman tensors for the CDO and FCO phases in the body-centered-tetragonal coordinate system.}	
	
\end{table*}

\subsection{Cell-doubled-orthorhombic phase}

\begin{figure}[htb]
	\centering
	\includegraphics[width = 1.0\linewidth]{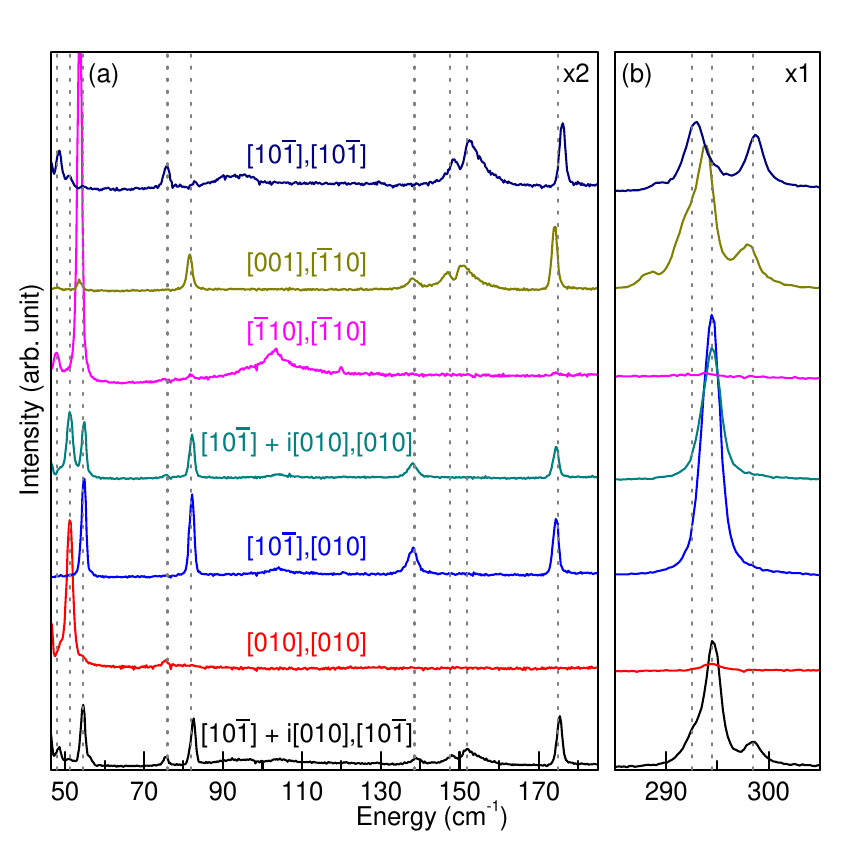}
	\caption{\label{doubled_polarization_supp} (Color online) Raman scattering spectra of Mn$_3$O$_4$ in the CDO phase at $T = 8 $~K in the energy ranges (a)~$50-185 \textrm{ cm}^{-1}$ and (b)~$285-305 \textrm{ cm}^{-1}$ taken with various polarizations of the incident ([hkl]) and scattered light ([h'k'l']), as indicated by [hkl],[h'k'l']. ($[h_1k_1l_1] + i[h_2k_2l_2]$ denotes circular polarization.) Vertical, dashed lines indicate magnon and phonon peak positions in the CDO phase. The data have been offset for clarity.}
\end{figure}

Seven configurations of incident and scattered light polarizations were used to determine the symmetries of excitations present in the CDO phase of Mn$_3$O$_4$.
Each sample was field-trained to create a single domain prior to measurement.
The obtained spectra are shown in Fig.~\ref{doubled_polarization_supp}.
For each polarization configuration, the scattering amplitudes of the irreducible representations of the $D_{2h}$ point group are tabulated in Table~\ref{table:CDOamplitudes_supp}.

Each CDO normal mode can be assigned to an irreducible representation of the $D_{2h}$ point group using the CDO Raman tensors presented in Table~\ref{table:tensors}.
This is \emph{not} possible using the FCO Raman tensors, indicating that the high-symmetry directions in the CDO phase are along the $[100]$ and $[010]$ directions, as described previously.
The experimentally determined energies and symmetries of the Raman-active magnetic and vibrational excitations with $\omega \leq 300 \text{ cm}^{-1}$ are tabulated in Table~\ref{table:CDOsymmetries}.
We note that the Raman tensor of the $\omega = 51 \text{ cm}^{-1}$ $A_g$ mode appears to have $a \simeq -b$, suggesting that this normal mode originates from a $B_{1g}$ mode in the tetragonal ($D_{4h}$) phase.

\begin{table}[h!]
	\centering
	\setlength{\tabcolsep}{5pt}
	\resizebox{\linewidth}{!}{
	\begin{tabular}{|c|c|c|c|c|c|}
		\hline
		\multicolumn{2}{|c|}{Polarization}	&	\multicolumn{4}{|c|}{CDO Amplitude ($|\mathbf{\hat{E}_s} \chi \mathbf{\hat{E}_i}|^2$)}	\\		\hline
		Incident ($\mathbf{E_i}$)   &  Scattered ($\mathbf{E_s}$)  &  $A_g$	&	$B_{1g}$	&	$B_{2g}$	&	$B_{3g}$  \\ \hline 
		$[10\bar{1}]$   &  $[10\bar{1}]$   &   $0.1a^2 + 0.4 ac + 0.5 c^2$	&	& $0.8e^2$	&   \\ \hline
		$[001]$  &   $[\bar{1}10]$   &	&	&   $0.5e^2$	& $0.5f^2$   \\ \hline
		$[\bar{1}10]$  &   $[\bar{1}10]$  &   $0.5(a + b)^2$	& $d^2$	&	&	  \\ \hline
		$[10\bar{1}] + i[010]$  &   $[010]$   &   $0.5b^2$	& $0.1d^2$	&	&$0.4f^2$  \\ \hline
		$[10\bar{1}]$  &   $[010]$  &	&	$0.3d^2$	&	&	$0.7f^2$   \\ \hline
		$[010]$  &   $[010]$  &   $b^2$	&	&	&  \\ \hline
		$[10\bar{1}] + i[010]$  &   $[10\bar{1}]$   &   $0.2ac + 0.3c^2$	&	$0.1d^2$	&	$0.4e^2$	&	$0.4f^2$  \\ \hline

	\end{tabular}
	}
	\caption{\label{table:CDOamplitudes_supp}Incident and scattered polarizations with expected amplitudes for each scattering geometry. ($[h_1k_1l_1] + i[h_2k_2l_2]$ denotes circular polarization.)}
	
\end{table}

\begin{table}[htb]
	\centering
	\begin{tabular}{|c|c|c|}
		\hline
		\multicolumn{3}{|c|}{CDO Excitation Symmetries}  \\ \hline
		\multirow{1}{*}{Excitation}   &   Energy (cm$^{-1}$)   &   $\Gamma$ ($D_{2h}$)  \\ \cline{1-3}
		\multicolumn{1}{ |c | }{\multirow{9}{*}{Magnon}}
		&  49   &   A$_{g}$   \\ \cline{2-3}
		&  51   &   A$_{g}$   \\ \cline{2-3}
		&  54   &   B$_{1g}$  \\ \cline{2-3}
		&  75   &   A$_{g}$   \\ \cline{2-3}
		&  83   &   B$_{3g}$  \\ \cline{2-3}
		&  139  &   B$_{3g}$  \\ \cline{2-3}
		&  148  &   B$_{2g}$  \\ \cline{2-3}
		&  152  &   B$_{2g}$  \\ \hline
		\multicolumn{1}{|c|}{\multirow{6}{*}{Phonon}}  
		&  175  &   B$_{2g}$  \\ \cline{2-3}
		&  175  &   B$_{3g}$  \\ \cline{2-3}
		&  293  &   B$_{2g}$  \\ \cline{2-3}
		&  295  &   B$_{3g}$  \\ \cline{2-3}
		&  298  &   B$_{2g}$  \\ \cline{1-3}
	\end{tabular}
	
	\caption{\label{table:CDOsymmetries}Energies and symmetries of the magnetic and vibrational excitations in the CDO phase at $T = 8$~K. The energy difference between the $B_{2g}$ and $B_{3g}$ phonons at $\omega = 175$~cm$^{-1}$ could not be resolved in this phase.}
	
\end{table}

\subsection{Face-centered-orthorhombic phase}

\begin{figure}[htb]
	\centering
	\includegraphics[width = 1.0\linewidth]{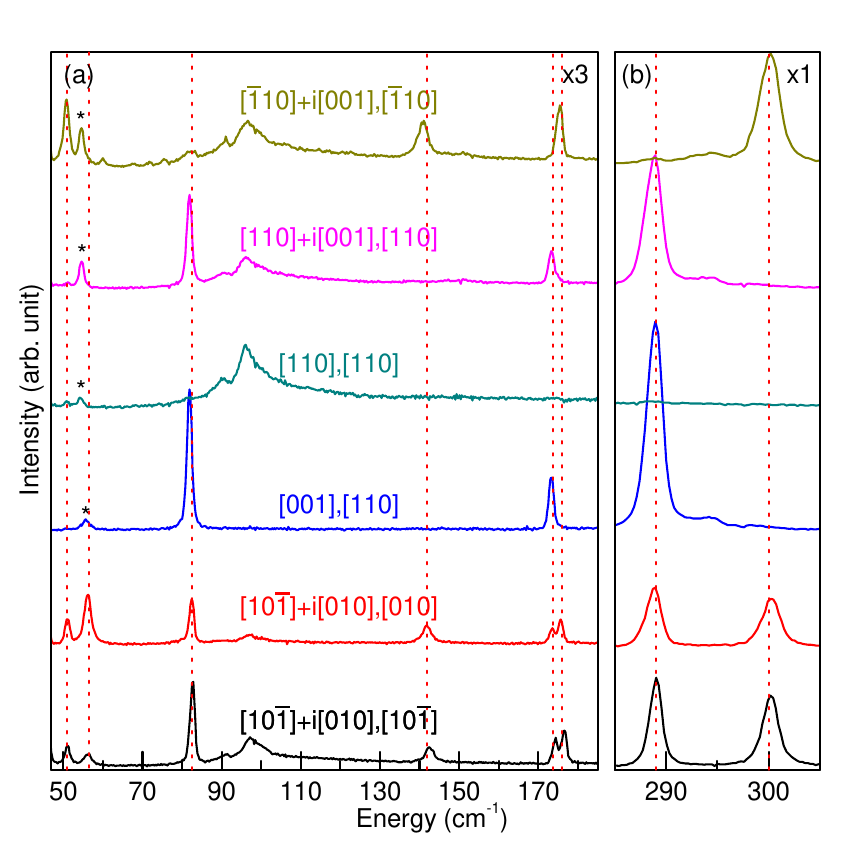}
	\caption{\label{undoubled_polarization_supp} (Color online) Raman scattering spectra of Mn$_3$O$_4$ in the FCO phase at $T = 8 $~K in the energy ranges (a)~$50-185 \textrm{ cm}^{-1}$ and (b)~$285-305 \textrm{ cm}^{-1}$ taken with various polarizations of the incident ([hkl]) and scattered light ([h'k'l']), as indicated by [hkl],[h'k'l']. ($[h_1k_1l_1] + i[h_2k_2l_2]$ denotes circular polarization.) Vertical, dashed lines indicate magnon and phonon peak positions in the FCO phase. The asterisk indicates a peak associated with a remnant CDO phase. The data have been offset for clarity.}
\end{figure}

Six configurations of incident and scattered light polarizations were used to determine the symmetries of excitations present in the FCO phase of Mn$_3$O$_4$.
Again, each sample was field-trained to create a single domain prior to measurement.
The obtained spectra are shown in Fig.~\ref{undoubled_polarization_supp}.
For each polarization configuration, the scattering amplitudes of the irreducible representations of the $D_{2h}$ point group are tabulated in Table~\ref{table:FCOamplitudes_supp}.

Each FCO normal mode can be assigned to an irreducible representation of the $D_{2h}$ point group using the FCO Raman tensors presented in Table~\ref{table:tensors}.
This is \emph{not} possible using the CDO Raman tensors, indicating that the high-symmetry directions in the FCO phase are along the $[\bar{1}10]$ and $[110]$ directions, as described previously.
The experimentally determined energies and symmetries of the Raman-active magnetic and vibrational excitations with $\omega \leq 300 \text{ cm}^{-1}$ are tabulated in Table~\ref{table:FCOsymmetries}.
The broad, asymmetric peak at $\omega = 96 \text{ cm}^{-1}$ in the FCO phase is identified as a two-magnon excitation based on the analysis of Gleason et al.~\cite{Gleason2014} and will not be discussed.

\begin{table}[htb]
	\centering
	\setlength{\tabcolsep}{5pt}
	\resizebox{\linewidth}{!}{
	\begin{tabular}{|c|c|c|c|c|c|}
		\hline
		\multicolumn{2}{|c|}{Polarization}	&	\multicolumn{4}{|c|}{FCO Amplitude ($|\mathbf{\hat{E}_s} \chi \mathbf{\hat{E}_i}|^2$)}	\\
		\hline
		Incident ($\mathbf{E_i}$)   &  Scattered ($\mathbf{E_s}$)  &	$A_g$	&	$B_{1g}$	&	$B_{2g}$	&	$B_{3g}$  \\ \hline 
		$[\bar{1}10] + i[001]$   &  $[\bar{1}10]$   &   $0.5a^2$	&	&	$0.5e^2$	&	   \\ \hline
		$[110] + i[001]$  &   $[001]$   &   $0.5b^2$	&	&	&	$0.5f^2$   \\ \hline
		$[110]$  &   $[110]$  &   $b^2$	&	&	&  \\ \hline
		$[001]$  &   $[110]$   &   &	&	&	$f^2$  \\ \hline
		$[10\bar{1}] + i[010]$  &   $[010]$  &   $0.2a^2 + 0.2ab +0.2b^2$	&	$0.5d^2$	&	$0.2e^2$	&	$0.2f^2$   \\ \hline
		$[10\bar{1}] + i[010]$  &   $[10\bar{1}]$   &   $0.1ac + 0.1bc + 0.3c^2$	&	&	$0.4e^2$	&	$0.4f^2$  \\ \hline

	\end{tabular}
	}
	
	\caption{\label{table:FCOamplitudes_supp}Incident and scattered polarizations as well as the expected amplitudes for each scattering geometry. ($[h_1k_1l_1] + i[h_2k_2l_2]$ denotes circular polarization.)}
	
\end{table}

\begin{table}[b]
	\centering
	\begin{tabular}{|c|c|c|}
		\hline
		\multicolumn{3}{|c|}{FCO Excitation Symmetries}  \\ \hline
		\multirow{1}{*}{Excitation}   &   Energy (cm$^{-1}$)   &   $\Gamma$ ($D_{2h}$)  \\ \cline{1-3}
		\multicolumn{1}{ |c | }{\multirow{5}{*}{Magnon}}
		&  51   &   A$_{g}$   \\ \cline{2-3}
		&  56   &   B$_{1g}$  \\ \cline{2-3}
		&  83   &   B$_{3g}$  \\ \cline{2-3}
		&  142  &   B$_{2g}$  \\ \hline
		\multicolumn{1}{|c|}{\multirow{5}{*}{Phonon}}  
		&  174  &   B$_{3g}$  \\ \cline{2-3}
		&  176  &   B$_{2g}$  \\ \cline{2-3}
		&  289  &   B$_{3g}$  \\ \cline{2-3}
		&  300  &   B$_{2g}$  \\ \cline{1-3}
	\end{tabular}
	
	\caption{\label{table:FCOsymmetries}Energies and symmetries of the magnetic and vibrational excitations in the FCO phase at $T = 8$~K.}
	
\end{table}

Determining the symmetry of the $\omega = 56$~cm$^{-1}$ magnetic excitation is complicated by the peak at $\omega = 54 \text{ cm}^{-1}$, which is a peak exclusive to the Raman spectrum of the CDO phase.
To see this, note that the $\omega = 54 \text{ cm}^{-1}$ peak has a large scattering cross-section, as seen in Fig.~\ref{doubled_polarization_supp}.
The nonzero peak intensity at $\omega = 54 \text{ cm}^{-1}$ in the FCO phase (Fig.~\ref{undoubled_polarization_supp}) is due to the presence of a small, remnant CDO phase fraction indicated by a nonzero peak intensity at $\omega = 295 \text{ cm}^{-1}$.
In our analysis, we conclude that the peak at $\omega = 56 \text{ cm}^{-1}$ is absent in all but the bottom two curves in Fig.~\ref{undoubled_polarization_supp}.

%\bibliography{mn3o4_refs}

\begin{thebibliography}{42}%
\makeatletter
\providecommand \@ifxundefined [1]{%
 \@ifx{#1\undefined}
}%
\providecommand \@ifnum [1]{%
 \ifnum #1\expandafter \@firstoftwo
 \else \expandafter \@secondoftwo
 \fi
}%
\providecommand \@ifx [1]{%
 \ifx #1\expandafter \@firstoftwo
 \else \expandafter \@secondoftwo
 \fi
}%
\providecommand \natexlab [1]{#1}%
\providecommand \enquote  [1]{``#1''}%
\providecommand \bibnamefont  [1]{#1}%
\providecommand \bibfnamefont [1]{#1}%
\providecommand \citenamefont [1]{#1}%
\providecommand \href@noop [0]{\@secondoftwo}%
\providecommand \href [0]{\begingroup \@sanitize@url \@href}%
\providecommand \@href[1]{\@@startlink{#1}\@@href}%
\providecommand \@@href[1]{\endgroup#1\@@endlink}%
\providecommand \@sanitize@url [0]{\catcode `\\12\catcode `\$12\catcode
  `\&12\catcode `\#12\catcode `\^12\catcode `\_12\catcode `\%12\relax}%
\providecommand \@@startlink[1]{}%
\providecommand \@@endlink[0]{}%
\providecommand \url  [0]{\begingroup\@sanitize@url \@url }%
\providecommand \@url [1]{\endgroup\@href {#1}{\urlprefix }}%
\providecommand \urlprefix  [0]{URL }%
\providecommand \Eprint [0]{\href }%
\providecommand \doibase [0]{http://dx.doi.org/}%
\providecommand \selectlanguage [0]{\@gobble}%
\providecommand \bibinfo  [0]{\@secondoftwo}%
\providecommand \bibfield  [0]{\@secondoftwo}%
\providecommand \translation [1]{[#1]}%
\providecommand \BibitemOpen [0]{}%
\providecommand \bibitemStop [0]{}%
\providecommand \bibitemNoStop [0]{.\EOS\space}%
\providecommand \EOS [0]{\spacefactor3000\relax}%
\providecommand \BibitemShut  [1]{\csname bibitem#1\endcsname}%
\let\auto@bib@innerbib\@empty
%</preamble>
\bibitem [{\citenamefont {Dagotto}(2005)}]{Dagotto2005}%
  \BibitemOpen
  \bibfield  {author} {\bibinfo {author} {\bibfnamefont {E.}~\bibnamefont
  {Dagotto}},\ }\href {\doibase 10.1126/science.1107559} {\bibfield  {journal}
  {\bibinfo  {journal} {Science}\ }\textbf {\bibinfo {volume} {309}},\ \bibinfo
  {pages} {257} (\bibinfo {year} {2005})}\BibitemShut {NoStop}%
\bibitem [{\citenamefont {Dagotto}\ \emph {et~al.}(2008)\citenamefont
  {Dagotto}, \citenamefont {Yunoki}, \citenamefont {\c{S}en}, \citenamefont
  {Alvarez},\ and\ \citenamefont {Moreo}}]{Dagotto2008}%
  \BibitemOpen
  \bibfield  {author} {\bibinfo {author} {\bibfnamefont {E.}~\bibnamefont
  {Dagotto}}, \bibinfo {author} {\bibfnamefont {S.}~\bibnamefont {Yunoki}},
  \bibinfo {author} {\bibfnamefont {C.}~\bibnamefont {\c{S}en}}, \bibinfo
  {author} {\bibfnamefont {G.}~\bibnamefont {Alvarez}}, \ and\ \bibinfo
  {author} {\bibfnamefont {A.}~\bibnamefont {Moreo}},\ }\href
  {http://stacks.iop.org/0953-8984/20/i=43/a=434224} {\bibfield  {journal}
  {\bibinfo  {journal} {Journal of Physics: Condensed Matter}\ }\textbf
  {\bibinfo {volume} {20}},\ \bibinfo {pages} {434224} (\bibinfo {year}
  {2008})}\BibitemShut {NoStop}%
\bibitem [{\citenamefont {Ahn}\ \emph {et~al.}(2013)\citenamefont {Ahn},
  \citenamefont {Seman}, \citenamefont {Lookman},\ and\ \citenamefont
  {Bishop}}]{Ahn2013}%
  \BibitemOpen
  \bibfield  {author} {\bibinfo {author} {\bibfnamefont {K.~H.}\ \bibnamefont
  {Ahn}}, \bibinfo {author} {\bibfnamefont {T.~F.}\ \bibnamefont {Seman}},
  \bibinfo {author} {\bibfnamefont {T.}~\bibnamefont {Lookman}}, \ and\
  \bibinfo {author} {\bibfnamefont {A.~R.}\ \bibnamefont {Bishop}},\ }\href
  {\doibase 10.1103/PhysRevB.88.144415} {\bibfield  {journal} {\bibinfo
  {journal} {Phys. Rev. B}\ }\textbf {\bibinfo {volume} {88}},\ \bibinfo
  {pages} {144415} (\bibinfo {year} {2013})}\BibitemShut {NoStop}%
\bibitem [{\citenamefont {Wu}\ and\ \citenamefont {Leighton}(2003)}]{Wu2003}%
  \BibitemOpen
  \bibfield  {author} {\bibinfo {author} {\bibfnamefont {J.}~\bibnamefont
  {Wu}}\ and\ \bibinfo {author} {\bibfnamefont {C.}~\bibnamefont {Leighton}},\
  }\href {\doibase 10.1103/PhysRevB.67.174408} {\bibfield  {journal} {\bibinfo
  {journal} {Phys. Rev. B}\ }\textbf {\bibinfo {volume} {67}},\ \bibinfo
  {pages} {174408} (\bibinfo {year} {2003})}\BibitemShut {NoStop}%
\bibitem [{\citenamefont {Snow}\ \emph {et~al.}(2001)\citenamefont {Snow},
  \citenamefont {Cooper}, \citenamefont {Young}, \citenamefont {Fisk},
  \citenamefont {Comment},\ and\ \citenamefont {Ansermet}}]{Snow2001}%
  \BibitemOpen
  \bibfield  {author} {\bibinfo {author} {\bibfnamefont {C.~S.}\ \bibnamefont
  {Snow}}, \bibinfo {author} {\bibfnamefont {S.~L.}\ \bibnamefont {Cooper}},
  \bibinfo {author} {\bibfnamefont {D.~P.}\ \bibnamefont {Young}}, \bibinfo
  {author} {\bibfnamefont {Z.}~\bibnamefont {Fisk}}, \bibinfo {author}
  {\bibfnamefont {A.}~\bibnamefont {Comment}}, \ and\ \bibinfo {author}
  {\bibfnamefont {J.-P.}\ \bibnamefont {Ansermet}},\ }\href {\doibase
  10.1103/PhysRevB.64.174412} {\bibfield  {journal} {\bibinfo  {journal} {Phys.
  Rev. B}\ }\textbf {\bibinfo {volume} {64}},\ \bibinfo {pages} {174412}
  (\bibinfo {year} {2001})}\BibitemShut {NoStop}%
\bibitem [{\citenamefont {{Kivelson}}\ \emph {et~al.}(1998)\citenamefont
  {{Kivelson}}, \citenamefont {{Fradkin}},\ and\ \citenamefont
  {{Emery}}}]{Kivelson1998}%
  \BibitemOpen
  \bibfield  {author} {\bibinfo {author} {\bibfnamefont {S.~A.}\ \bibnamefont
  {{Kivelson}}}, \bibinfo {author} {\bibfnamefont {E.}~\bibnamefont
  {{Fradkin}}}, \ and\ \bibinfo {author} {\bibfnamefont {V.~J.}\ \bibnamefont
  {{Emery}}},\ }\href {\doibase 10.1038/31177} {\bibfield  {journal} {\bibinfo
  {journal} {\nat}\ }\textbf {\bibinfo {volume} {393}},\ \bibinfo {pages} {550}
  (\bibinfo {year} {1998})}\BibitemShut {NoStop}%
\bibitem [{\citenamefont {Kimura}\ \emph {et~al.}(2003)\citenamefont {Kimura},
  \citenamefont {Goto}, \citenamefont {Shintani}, \citenamefont {Ishizaka},
  \citenamefont {Arima},\ and\ \citenamefont {Tokura}}]{Kimura2003}%
  \BibitemOpen
  \bibfield  {author} {\bibinfo {author} {\bibfnamefont {T.}~\bibnamefont
  {Kimura}}, \bibinfo {author} {\bibfnamefont {T.}~\bibnamefont {Goto}},
  \bibinfo {author} {\bibfnamefont {H.}~\bibnamefont {Shintani}}, \bibinfo
  {author} {\bibfnamefont {K.}~\bibnamefont {Ishizaka}}, \bibinfo {author}
  {\bibfnamefont {T.}~\bibnamefont {Arima}}, \ and\ \bibinfo {author}
  {\bibfnamefont {Y.}~\bibnamefont {Tokura}},\ }\href {\doibase
  10.1038/nature02018} {\bibfield  {journal} {\bibinfo  {journal} {Nature}\
  }\textbf {\bibinfo {volume} {426}},\ \bibinfo {pages} {55} (\bibinfo {year}
  {2003})}\BibitemShut {NoStop}%
\bibitem [{\citenamefont {Ramirez}(1997)}]{Ramirez1997}%
  \BibitemOpen
  \bibfield  {author} {\bibinfo {author} {\bibfnamefont {A.~P.}\ \bibnamefont
  {Ramirez}},\ }\href {http://stacks.iop.org/0953-8984/9/i=39/a=005} {\bibfield
   {journal} {\bibinfo  {journal} {Journal of Physics: Condensed Matter}\
  }\textbf {\bibinfo {volume} {9}},\ \bibinfo {pages} {8171} (\bibinfo {year}
  {1997})}\BibitemShut {NoStop}%
\bibitem [{\citenamefont {Lavrov}\ \emph {et~al.}(2002)\citenamefont {Lavrov},
  \citenamefont {Komiya},\ and\ \citenamefont {Ando}}]{Lavrov2002}%
  \BibitemOpen
  \bibfield  {author} {\bibinfo {author} {\bibfnamefont {A.~N.}\ \bibnamefont
  {Lavrov}}, \bibinfo {author} {\bibfnamefont {S.}~\bibnamefont {Komiya}}, \
  and\ \bibinfo {author} {\bibfnamefont {Y.}~\bibnamefont {Ando}},\ }\href
  {\doibase 10.1038/418385a} {\bibfield  {journal} {\bibinfo  {journal}
  {Nature}\ }\textbf {\bibinfo {volume} {418}},\ \bibinfo {pages} {385}
  (\bibinfo {year} {2002})}\BibitemShut {NoStop}%
\bibitem [{\citenamefont {Kim}\ \emph {et~al.}(2010)\citenamefont {Kim},
  \citenamefont {Chen}, \citenamefont {Joe}, \citenamefont {Fradkin},
  \citenamefont {Abbamonte},\ and\ \citenamefont {Cooper}}]{Kim2010}%
  \BibitemOpen
  \bibfield  {author} {\bibinfo {author} {\bibfnamefont {M.}~\bibnamefont
  {Kim}}, \bibinfo {author} {\bibfnamefont {X.~M.}\ \bibnamefont {Chen}},
  \bibinfo {author} {\bibfnamefont {Y.~I.}\ \bibnamefont {Joe}}, \bibinfo
  {author} {\bibfnamefont {E.}~\bibnamefont {Fradkin}}, \bibinfo {author}
  {\bibfnamefont {P.}~\bibnamefont {Abbamonte}}, \ and\ \bibinfo {author}
  {\bibfnamefont {S.~L.}\ \bibnamefont {Cooper}},\ }\href {\doibase
  10.1103/PhysRevLett.104.136402} {\bibfield  {journal} {\bibinfo  {journal}
  {Phys. Rev. Lett.}\ }\textbf {\bibinfo {volume} {104}},\ \bibinfo {pages}
  {136402} (\bibinfo {year} {2010})}\BibitemShut {NoStop}%
\bibitem [{\citenamefont {Lawes}\ \emph {et~al.}(2003)\citenamefont {Lawes},
  \citenamefont {Ramirez}, \citenamefont {Varma},\ and\ \citenamefont
  {Subramanian}}]{Lawes2003}%
  \BibitemOpen
  \bibfield  {author} {\bibinfo {author} {\bibfnamefont {G.}~\bibnamefont
  {Lawes}}, \bibinfo {author} {\bibfnamefont {A.~P.}\ \bibnamefont {Ramirez}},
  \bibinfo {author} {\bibfnamefont {C.~M.}\ \bibnamefont {Varma}}, \ and\
  \bibinfo {author} {\bibfnamefont {M.~A.}\ \bibnamefont {Subramanian}},\
  }\href {\doibase 10.1103/PhysRevLett.91.257208} {\bibfield  {journal}
  {\bibinfo  {journal} {Phys. Rev. Lett.}\ }\textbf {\bibinfo {volume} {91}},\
  \bibinfo {pages} {257208} (\bibinfo {year} {2003})}\BibitemShut {NoStop}%
\bibitem [{\citenamefont {Gschneidner}\ \emph {et~al.}(2005)\citenamefont
  {Gschneidner}, \citenamefont {Pecharsky},\ and\ \citenamefont
  {Tsokol}}]{Gschneidner2005}%
  \BibitemOpen
  \bibfield  {author} {\bibinfo {author} {\bibfnamefont {K.~A.}\ \bibnamefont
  {Gschneidner}, \bibfnamefont {Jr.}}, \bibinfo {author} {\bibfnamefont
  {V.~K.}\ \bibnamefont {Pecharsky}}, \ and\ \bibinfo {author} {\bibfnamefont
  {A.~O.}\ \bibnamefont {Tsokol}},\ }\href {\doibase
  10.1088/0034-4885/68/6/R04} {\bibfield  {journal} {\bibinfo  {journal}
  {Reports on Progress in Physics}\ }\textbf {\bibinfo {volume} {68}},\
  \bibinfo {pages} {1479} (\bibinfo {year} {2005})}\BibitemShut {NoStop}%
\bibitem [{\citenamefont {Tackett}\ \emph {et~al.}(2007)\citenamefont
  {Tackett}, \citenamefont {Lawes}, \citenamefont {Melot}, \citenamefont
  {Grossman}, \citenamefont {Toberer},\ and\ \citenamefont
  {Seshadri}}]{Tackett2007}%
  \BibitemOpen
  \bibfield  {author} {\bibinfo {author} {\bibfnamefont {R.}~\bibnamefont
  {Tackett}}, \bibinfo {author} {\bibfnamefont {G.}~\bibnamefont {Lawes}},
  \bibinfo {author} {\bibfnamefont {B.~C.}\ \bibnamefont {Melot}}, \bibinfo
  {author} {\bibfnamefont {M.}~\bibnamefont {Grossman}}, \bibinfo {author}
  {\bibfnamefont {E.~S.}\ \bibnamefont {Toberer}}, \ and\ \bibinfo {author}
  {\bibfnamefont {R.}~\bibnamefont {Seshadri}},\ }\href {\doibase
  10.1103/PhysRevB.76.024409} {\bibfield  {journal} {\bibinfo  {journal}
  {Physical Review B}\ }\textbf {\bibinfo {volume} {76}},\ \bibinfo {pages}
  {024409} (\bibinfo {year} {2007})}\BibitemShut {NoStop}%
\bibitem [{\citenamefont {Lacroix}\ \emph {et~al.}(2011)\citenamefont
  {Lacroix}, \citenamefont {Mendels},\ and\ \citenamefont
  {Mila}}]{Lacroix2011}%
  \BibitemOpen
  \bibfield  {author} {\bibinfo {author} {\bibfnamefont {C.}~\bibnamefont
  {Lacroix}}, \bibinfo {author} {\bibfnamefont {P.}~\bibnamefont {Mendels}}, \
  and\ \bibinfo {author} {\bibfnamefont {F.}~\bibnamefont {Mila}},\ }\href
  {http://scholar.google.com/scholar?hl=en{\&}btnG=Search{\&}q=intitle:Introduction+to+Frustrated+Magnetism{\#}1
  http://books.google.com/books?hl=en{\&}lr={\&}id=utSV09ZuhOkC{\&}oi=fnd{\&}pg=PR5{\&}dq=Introduction+to+Frustrated+Magnetism:+Materials,+Experiments,+Theory{\&}ots=25bYLuZM}
  {\emph {\bibinfo {title} {{Introduction to Frustrated Magnetism: Materials,
  Experiments, Theory}}}}\ (\bibinfo {year} {2011})\BibitemShut {NoStop}%
\bibitem [{\citenamefont {McCallum}\ \emph {et~al.}(1976)\citenamefont
  {McCallum}, \citenamefont {Johnston}, \citenamefont {Luengo},\ and\
  \citenamefont {Maple}}]{McCallum1976}%
  \BibitemOpen
  \bibfield  {author} {\bibinfo {author} {\bibfnamefont {R.~W.}\ \bibnamefont
  {McCallum}}, \bibinfo {author} {\bibfnamefont {D.~C.}\ \bibnamefont
  {Johnston}}, \bibinfo {author} {\bibfnamefont {C.~A.}\ \bibnamefont
  {Luengo}}, \ and\ \bibinfo {author} {\bibfnamefont {M.~B.}\ \bibnamefont
  {Maple}},\ }\href {\doibase 10.1007/BF00654828} {\bibfield  {journal}
  {\bibinfo  {journal} {Journal of Low Temperature Physics}\ }\textbf {\bibinfo
  {volume} {25}},\ \bibinfo {pages} {177} (\bibinfo {year} {1976})}\BibitemShut
  {NoStop}%
\bibitem [{\citenamefont {Irizawa}\ \emph {et~al.}(2011)\citenamefont
  {Irizawa}, \citenamefont {Suga}, \citenamefont {Isoyama}, \citenamefont
  {Shimai}, \citenamefont {Sato}, \citenamefont {Iizuka}, \citenamefont
  {Nanba}, \citenamefont {Higashiya}, \citenamefont {Niitaka},\ and\
  \citenamefont {Takagi}}]{Irizawa2011}%
  \BibitemOpen
  \bibfield  {author} {\bibinfo {author} {\bibfnamefont {A.}~\bibnamefont
  {Irizawa}}, \bibinfo {author} {\bibfnamefont {S.}~\bibnamefont {Suga}},
  \bibinfo {author} {\bibfnamefont {G.}~\bibnamefont {Isoyama}}, \bibinfo
  {author} {\bibfnamefont {K.}~\bibnamefont {Shimai}}, \bibinfo {author}
  {\bibfnamefont {K.}~\bibnamefont {Sato}}, \bibinfo {author} {\bibfnamefont
  {K.}~\bibnamefont {Iizuka}}, \bibinfo {author} {\bibfnamefont
  {T.}~\bibnamefont {Nanba}}, \bibinfo {author} {\bibfnamefont
  {A.}~\bibnamefont {Higashiya}}, \bibinfo {author} {\bibfnamefont
  {S.}~\bibnamefont {Niitaka}}, \ and\ \bibinfo {author} {\bibfnamefont
  {H.}~\bibnamefont {Takagi}},\ }\href {\doibase 10.1103/PhysRevB.84.235116}
  {\bibfield  {journal} {\bibinfo  {journal} {Phys. Rev. B}\ }\textbf {\bibinfo
  {volume} {84}},\ \bibinfo {pages} {235116} (\bibinfo {year}
  {2011})}\BibitemShut {NoStop}%
\bibitem [{\citenamefont {Kope\'{c}}\ \emph {et~al.}(2009)\citenamefont
  {Kope\'{c}}, \citenamefont {Dygas}, \citenamefont {Krok}, \citenamefont
  {Mauger}, \citenamefont {Gendron}, \citenamefont {Jaszczak-Figiel},
  \citenamefont {Gagor}, \citenamefont {Zaghib},\ and\ \citenamefont
  {Julien}}]{Kopec2009}%
  \BibitemOpen
  \bibfield  {author} {\bibinfo {author} {\bibfnamefont {M.}~\bibnamefont
  {Kope\'{c}}}, \bibinfo {author} {\bibfnamefont {J.~R.}\ \bibnamefont
  {Dygas}}, \bibinfo {author} {\bibfnamefont {F.}~\bibnamefont {Krok}},
  \bibinfo {author} {\bibfnamefont {A.}~\bibnamefont {Mauger}}, \bibinfo
  {author} {\bibfnamefont {F.}~\bibnamefont {Gendron}}, \bibinfo {author}
  {\bibfnamefont {B.}~\bibnamefont {Jaszczak-Figiel}}, \bibinfo {author}
  {\bibfnamefont {A.}~\bibnamefont {Gagor}}, \bibinfo {author} {\bibfnamefont
  {K.}~\bibnamefont {Zaghib}}, \ and\ \bibinfo {author} {\bibfnamefont {C.~M.}\
  \bibnamefont {Julien}},\ }\href {\doibase 10.1021/cm900609n} {\bibfield
  {journal} {\bibinfo  {journal} {Chemistry of Materials}\ }\textbf {\bibinfo
  {volume} {21}},\ \bibinfo {pages} {2525} (\bibinfo {year}
  {2009})}\BibitemShut {NoStop}%
\bibitem [{\citenamefont {Yamasaki}\ \emph {et~al.}(2006)\citenamefont
  {Yamasaki}, \citenamefont {Miyasaka}, \citenamefont {Kaneko}, \citenamefont
  {He}, \citenamefont {Arima},\ and\ \citenamefont {Tokura}}]{Yamasaki2006}%
  \BibitemOpen
  \bibfield  {author} {\bibinfo {author} {\bibfnamefont {Y.}~\bibnamefont
  {Yamasaki}}, \bibinfo {author} {\bibfnamefont {S.}~\bibnamefont {Miyasaka}},
  \bibinfo {author} {\bibfnamefont {Y.}~\bibnamefont {Kaneko}}, \bibinfo
  {author} {\bibfnamefont {J.-P.}\ \bibnamefont {He}}, \bibinfo {author}
  {\bibfnamefont {T.}~\bibnamefont {Arima}}, \ and\ \bibinfo {author}
  {\bibfnamefont {Y.}~\bibnamefont {Tokura}},\ }\href {\doibase
  10.1103/PhysRevLett.96.207204} {\bibfield  {journal} {\bibinfo  {journal}
  {Physical Review Letters}\ }\textbf {\bibinfo {volume} {96}},\ \bibinfo
  {pages} {207204} (\bibinfo {year} {2006})}\BibitemShut {NoStop}%
\bibitem [{\citenamefont {Dey}\ \emph {et~al.}(2014)\citenamefont {Dey},
  \citenamefont {Majumdar},\ and\ \citenamefont {Giri}}]{Dey2014}%
  \BibitemOpen
  \bibfield  {author} {\bibinfo {author} {\bibfnamefont {K.}~\bibnamefont
  {Dey}}, \bibinfo {author} {\bibfnamefont {S.}~\bibnamefont {Majumdar}}, \
  and\ \bibinfo {author} {\bibfnamefont {S.}~\bibnamefont {Giri}},\ }\href
  {\doibase 10.1103/PhysRevB.90.184424} {\bibfield  {journal} {\bibinfo
  {journal} {Physical Review B}\ }\textbf {\bibinfo {volume} {90}},\ \bibinfo
  {pages} {184424} (\bibinfo {year} {2014})}\BibitemShut {NoStop}%
\bibitem [{\citenamefont {Giovannetti}\ \emph {et~al.}(2011)\citenamefont
  {Giovannetti}, \citenamefont {Stroppa}, \citenamefont {Picozzi},
  \citenamefont {Baldomir}, \citenamefont {Pardo}, \citenamefont
  {Blanco-Canosa}, \citenamefont {Rivadulla}, \citenamefont {Jodlauk},
  \citenamefont {Niermann}, \citenamefont {Rohrkamp}, \citenamefont {Lorenz},
  \citenamefont {Streltsov}, \citenamefont {Khomskii},\ and\ \citenamefont
  {Hemberger}}]{Giovannetti2011}%
  \BibitemOpen
  \bibfield  {author} {\bibinfo {author} {\bibfnamefont {G.}~\bibnamefont
  {Giovannetti}}, \bibinfo {author} {\bibfnamefont {A.}~\bibnamefont
  {Stroppa}}, \bibinfo {author} {\bibfnamefont {S.}~\bibnamefont {Picozzi}},
  \bibinfo {author} {\bibfnamefont {D.}~\bibnamefont {Baldomir}}, \bibinfo
  {author} {\bibfnamefont {V.}~\bibnamefont {Pardo}}, \bibinfo {author}
  {\bibfnamefont {S.}~\bibnamefont {Blanco-Canosa}}, \bibinfo {author}
  {\bibfnamefont {F.}~\bibnamefont {Rivadulla}}, \bibinfo {author}
  {\bibfnamefont {S.}~\bibnamefont {Jodlauk}}, \bibinfo {author} {\bibfnamefont
  {D.}~\bibnamefont {Niermann}}, \bibinfo {author} {\bibfnamefont
  {J.}~\bibnamefont {Rohrkamp}}, \bibinfo {author} {\bibfnamefont
  {T.}~\bibnamefont {Lorenz}}, \bibinfo {author} {\bibfnamefont
  {S.}~\bibnamefont {Streltsov}}, \bibinfo {author} {\bibfnamefont {D.~I.}\
  \bibnamefont {Khomskii}}, \ and\ \bibinfo {author} {\bibfnamefont
  {J.}~\bibnamefont {Hemberger}},\ }\href {\doibase 10.1103/PhysRevB.83.060402}
  {\bibfield  {journal} {\bibinfo  {journal} {Physical Review B}\ }\textbf
  {\bibinfo {volume} {83}},\ \bibinfo {pages} {060402} (\bibinfo {year}
  {2011})}\BibitemShut {NoStop}%
\bibitem [{\citenamefont {Maignan}\ \emph {et~al.}(2012)\citenamefont
  {Maignan}, \citenamefont {Martin}, \citenamefont {Singh}, \citenamefont
  {{{Ch. Simon}}}, \citenamefont {Lebedev},\ and\ \citenamefont
  {Turner}}]{Maignan2012}%
  \BibitemOpen
  \bibfield  {author} {\bibinfo {author} {\bibfnamefont {A.}~\bibnamefont
  {Maignan}}, \bibinfo {author} {\bibfnamefont {C.}~\bibnamefont {Martin}},
  \bibinfo {author} {\bibfnamefont {K.}~\bibnamefont {Singh}}, \bibinfo
  {author} {\bibnamefont {{{Ch. Simon}}}}, \bibinfo {author} {\bibfnamefont
  {O.~I.}\ \bibnamefont {Lebedev}}, \ and\ \bibinfo {author} {\bibfnamefont
  {S.}~\bibnamefont {Turner}},\ }\href {\doibase 10.1016/j.jssc.2012.01.063}
  {\bibfield  {journal} {\bibinfo  {journal} {Journal of Solid State
  Chemistry}\ }\textbf {\bibinfo {volume} {195}},\ \bibinfo {pages} {41}
  (\bibinfo {year} {2012})}\BibitemShut {NoStop}%
\bibitem [{\citenamefont {Singh}\ \emph {et~al.}(2011)\citenamefont {Singh},
  \citenamefont {Maignan}, \citenamefont {Simon},\ and\ \citenamefont
  {Martin}}]{Singh2011}%
  \BibitemOpen
  \bibfield  {author} {\bibinfo {author} {\bibfnamefont {K.}~\bibnamefont
  {Singh}}, \bibinfo {author} {\bibfnamefont {A.}~\bibnamefont {Maignan}},
  \bibinfo {author} {\bibfnamefont {C.}~\bibnamefont {Simon}}, \ and\ \bibinfo
  {author} {\bibfnamefont {C.}~\bibnamefont {Martin}},\ }\href {\doibase
  10.1063/1.3656711} {\bibfield  {journal} {\bibinfo  {journal} {Applied
  Physics Letters}\ }\textbf {\bibinfo {volume} {99}},\ \bibinfo {pages}
  {172903} (\bibinfo {year} {2011})}\BibitemShut {NoStop}%
\bibitem [{\citenamefont {Suzuki}\ and\ \citenamefont
  {Katsufuji}(2008)}]{Suzuki2008}%
  \BibitemOpen
  \bibfield  {author} {\bibinfo {author} {\bibfnamefont {T.}~\bibnamefont
  {Suzuki}}\ and\ \bibinfo {author} {\bibfnamefont {T.}~\bibnamefont
  {Katsufuji}},\ }\href {\doibase 10.1103/PhysRevB.77.220402} {\bibfield
  {journal} {\bibinfo  {journal} {Physical Review B}\ }\textbf {\bibinfo
  {volume} {77}},\ \bibinfo {pages} {220402} (\bibinfo {year}
  {2008})}\BibitemShut {NoStop}%
\bibitem [{\citenamefont {Nii}\ \emph {et~al.}(2013)\citenamefont {Nii},
  \citenamefont {Sagayama}, \citenamefont {Umetsu}, \citenamefont {Abe},
  \citenamefont {Taniguchi},\ and\ \citenamefont {Arima}}]{Nii2013}%
  \BibitemOpen
  \bibfield  {author} {\bibinfo {author} {\bibfnamefont {Y.}~\bibnamefont
  {Nii}}, \bibinfo {author} {\bibfnamefont {H.}~\bibnamefont {Sagayama}},
  \bibinfo {author} {\bibfnamefont {H.}~\bibnamefont {Umetsu}}, \bibinfo
  {author} {\bibfnamefont {N.}~\bibnamefont {Abe}}, \bibinfo {author}
  {\bibfnamefont {K.}~\bibnamefont {Taniguchi}}, \ and\ \bibinfo {author}
  {\bibfnamefont {T.}~\bibnamefont {Arima}},\ }\href {\doibase
  10.1103/PhysRevB.87.195115} {\bibfield  {journal} {\bibinfo  {journal} {Phys.
  Rev. B}\ }\textbf {\bibinfo {volume} {87}},\ \bibinfo {pages} {195115}
  (\bibinfo {year} {2013})}\BibitemShut {NoStop}%
\bibitem [{\citenamefont {Kemei}\ \emph {et~al.}(2014)\citenamefont {Kemei},
  \citenamefont {Harada}, \citenamefont {Seshadri},\ and\ \citenamefont
  {Suchomel}}]{Kemei2014c}%
  \BibitemOpen
  \bibfield  {author} {\bibinfo {author} {\bibfnamefont {M.~C.}\ \bibnamefont
  {Kemei}}, \bibinfo {author} {\bibfnamefont {J.~K.}\ \bibnamefont {Harada}},
  \bibinfo {author} {\bibfnamefont {R.}~\bibnamefont {Seshadri}}, \ and\
  \bibinfo {author} {\bibfnamefont {M.~R.}\ \bibnamefont {Suchomel}},\ }\href
  {\doibase 10.1103/PhysRevB.90.064418} {\bibfield  {journal} {\bibinfo
  {journal} {Physical Review B}\ }\textbf {\bibinfo {volume} {90}},\ \bibinfo
  {pages} {064418} (\bibinfo {year} {2014})}\BibitemShut {NoStop}%
\bibitem [{\citenamefont {Van~Hook}\ and\ \citenamefont
  {Keith}(1958)}]{VanHook1958}%
  \BibitemOpen
  \bibfield  {author} {\bibinfo {author} {\bibfnamefont {H.~J.}\ \bibnamefont
  {Van~Hook}}\ and\ \bibinfo {author} {\bibfnamefont {M.~L.}\ \bibnamefont
  {Keith}},\ }\href@noop {} {\bibfield  {journal} {\bibinfo  {journal} {Am.
  Mineral.}\ }\textbf {\bibinfo {volume} {43}},\ \bibinfo {pages} {69}
  (\bibinfo {year} {1958})}\BibitemShut {NoStop}%
\bibitem [{\citenamefont {Jensen}\ and\ \citenamefont
  {Nielsen}(1974)}]{Jensen1974}%
  \BibitemOpen
  \bibfield  {author} {\bibinfo {author} {\bibfnamefont {G.~B.}\ \bibnamefont
  {Jensen}}\ and\ \bibinfo {author} {\bibfnamefont {O.~V.}\ \bibnamefont
  {Nielsen}},\ }\href {http://stacks.iop.org/0022-3719/7/i=2/a=019} {\bibfield
  {journal} {\bibinfo  {journal} {Journal of Physics C: Solid State Physics}\
  }\textbf {\bibinfo {volume} {7}},\ \bibinfo {pages} {409} (\bibinfo {year}
  {1974})}\BibitemShut {NoStop}%
\bibitem [{\citenamefont {Chardon}\ and\ \citenamefont
  {Vigneron}(1986)}]{Chardon1986}%
  \BibitemOpen
  \bibfield  {author} {\bibinfo {author} {\bibfnamefont {B.}~\bibnamefont
  {Chardon}}\ and\ \bibinfo {author} {\bibfnamefont {F.}~\bibnamefont
  {Vigneron}},\ }\href {\doibase
  http://dx.doi.org/10.1016/0304-8853(86)90132-0} {\bibfield  {journal}
  {\bibinfo  {journal} {Journal of Magnetism and Magnetic Materials}\ }\textbf
  {\bibinfo {volume} {58}},\ \bibinfo {pages} {128 } (\bibinfo {year}
  {1986})}\BibitemShut {NoStop}%
\bibitem [{\citenamefont {Chung}\ \emph {et~al.}(2013)\citenamefont {Chung},
  \citenamefont {Lee}, \citenamefont {Song}, \citenamefont {Suzuki},\ and\
  \citenamefont {Katsufuji}}]{Chung2013b}%
  \BibitemOpen
  \bibfield  {author} {\bibinfo {author} {\bibfnamefont {J.~H.}\ \bibnamefont
  {Chung}}, \bibinfo {author} {\bibfnamefont {K.~H.}\ \bibnamefont {Lee}},
  \bibinfo {author} {\bibfnamefont {Y.~S.}\ \bibnamefont {Song}}, \bibinfo
  {author} {\bibfnamefont {T.}~\bibnamefont {Suzuki}}, \ and\ \bibinfo {author}
  {\bibfnamefont {T.}~\bibnamefont {Katsufuji}},\ }\href
  {http://scattering.korea.ac.kr/pdfs/JPSJ-82-034707.pdf} {\bibfield  {journal}
  {\bibinfo  {journal} {Journal of the Physical Society Japan}\ }\textbf
  {\bibinfo {volume} {82}},\ \bibinfo {pages} {034707} (\bibinfo {year}
  {2013})}\BibitemShut {NoStop}%
\bibitem [{\citenamefont {Kim}\ \emph {et~al.}(2011)\citenamefont {Kim},
  \citenamefont {Chen}, \citenamefont {Wang}, \citenamefont {Nelson},
  \citenamefont {Budakian}, \citenamefont {Abbamonte},\ and\ \citenamefont
  {Cooper}}]{Kim2011a}%
  \BibitemOpen
  \bibfield  {author} {\bibinfo {author} {\bibfnamefont {M.}~\bibnamefont
  {Kim}}, \bibinfo {author} {\bibfnamefont {X.~M.}\ \bibnamefont {Chen}},
  \bibinfo {author} {\bibfnamefont {X.}~\bibnamefont {Wang}}, \bibinfo {author}
  {\bibfnamefont {C.~S.}\ \bibnamefont {Nelson}}, \bibinfo {author}
  {\bibfnamefont {R.}~\bibnamefont {Budakian}}, \bibinfo {author}
  {\bibfnamefont {P.}~\bibnamefont {Abbamonte}}, \ and\ \bibinfo {author}
  {\bibfnamefont {S.~L.}\ \bibnamefont {Cooper}},\ }\href {\doibase
  10.1103/PhysRevB.84.174424} {\bibfield  {journal} {\bibinfo  {journal} {Phys.
  Rev. B}\ }\textbf {\bibinfo {volume} {84}},\ \bibinfo {pages} {174424}
  (\bibinfo {year} {2011})}\BibitemShut {NoStop}%
\bibitem [{\citenamefont {Nielsen}(1969)}]{Nielsen1969}%
  \BibitemOpen
  \bibfield  {author} {\bibinfo {author} {\bibfnamefont {O.~V.}\ \bibnamefont
  {Nielsen}},\ }\href
  {http://www.sciencedirect.com/science/article/pii/0022024869900426}
  {\bibfield  {journal} {\bibinfo  {journal} {Journal of Crystal Growth}\
  }\textbf {\bibinfo {volume} {5}},\ \bibinfo {pages} {398} (\bibinfo {year}
  {1969})}\BibitemShut {NoStop}%
\bibitem [{\citenamefont {Yamamoto}\ \emph {et~al.}(1972)\citenamefont
  {Yamamoto}, \citenamefont {Nagasawa}, \citenamefont {Bando},\ and\
  \citenamefont {Takada}}]{Yamamoto1972}%
  \BibitemOpen
  \bibfield  {author} {\bibinfo {author} {\bibfnamefont {N.}~\bibnamefont
  {Yamamoto}}, \bibinfo {author} {\bibfnamefont {K.}~\bibnamefont {Nagasawa}},
  \bibinfo {author} {\bibfnamefont {Y.}~\bibnamefont {Bando}}, \ and\ \bibinfo
  {author} {\bibfnamefont {T.}~\bibnamefont {Takada}},\ }\href@noop {}
  {\bibfield  {journal} {\bibinfo  {journal} {Japanese Journal of Applied
  Physics}\ }\textbf {\bibinfo {volume} {11}},\ \bibinfo {pages} {1754 }
  (\bibinfo {year} {1972})}\BibitemShut {NoStop}%
\bibitem [{\citenamefont {Wanklyn}(1981)}]{Wanklyn1981}%
  \BibitemOpen
  \bibfield  {author} {\bibinfo {author} {\bibfnamefont {B.~M.}\ \bibnamefont
  {Wanklyn}},\ }\href {\doibase 10.1016/0022-0248(81)90525-X} {\bibfield
  {journal} {\bibinfo  {journal} {Journal of Crystal Growth}\ }\textbf
  {\bibinfo {volume} {54}},\ \bibinfo {pages} {610} (\bibinfo {year}
  {1981})}\BibitemShut {NoStop}%
\bibitem [{\citenamefont {Couderc}\ \emph {et~al.}(1994)\citenamefont
  {Couderc}, \citenamefont {Fritsch}, \citenamefont {Brieu}, \citenamefont
  {Vanderschaeve}, \citenamefont {Fagot},\ and\ \citenamefont
  {Rousset}}]{Couderc1994}%
  \BibitemOpen
  \bibfield  {author} {\bibinfo {author} {\bibfnamefont {J.~J.}\ \bibnamefont
  {Couderc}}, \bibinfo {author} {\bibfnamefont {S.}~\bibnamefont {Fritsch}},
  \bibinfo {author} {\bibfnamefont {M.}~\bibnamefont {Brieu}}, \bibinfo
  {author} {\bibfnamefont {G.}~\bibnamefont {Vanderschaeve}}, \bibinfo {author}
  {\bibfnamefont {M.}~\bibnamefont {Fagot}}, \ and\ \bibinfo {author}
  {\bibfnamefont {A.}~\bibnamefont {Rousset}},\ }\href {\doibase
  10.1080/01418639408240274} {\bibfield  {journal} {\bibinfo  {journal}
  {Philosophical Magazine Part B}\ }\textbf {\bibinfo {volume} {70}},\ \bibinfo
  {pages} {1077} (\bibinfo {year} {1994})}\BibitemShut {NoStop}%
\bibitem [{\citenamefont {Kim}(2011)}]{Kim2011b}%
  \BibitemOpen
  \bibfield  {author} {\bibinfo {author} {\bibfnamefont {M.}~\bibnamefont
  {Kim}},\ }\emph {\bibinfo {title} {Magnetic field- and pressure-tuned phases
  in La$_{x}$Pr$_{y}$Ca$_{1-x-y}$MnO$_{3}$ and Mn$_{3}$O$_{4}$: Inelastic light
  scattering studies and single crystal growth}},\ \href@noop {} {Ph.D.
  thesis},\ \bibinfo  {school} {University of Illinois at Urbana-Champaign}
  (\bibinfo {year} {2011})\BibitemShut {NoStop}%
\bibitem [{dom()}]{domainalignment}%
  \BibitemOpen
  \href@noop {} {}\bibinfo {note} {A magnetic field sufficiently large to
  remove any signature of domains in a polarization study. In the present case,
  approximately 0.2~T.}\BibitemShut {Stop}%
\bibitem [{\citenamefont {Takubo}\ \emph {et~al.}(2011)\citenamefont {Takubo},
  \citenamefont {Kubota}, \citenamefont {Suzuki}, \citenamefont {Kanzaki},
  \citenamefont {Miyahara}, \citenamefont {Furukawa},\ and\ \citenamefont
  {Katsufuji}}]{Takubo2011}%
  \BibitemOpen
  \bibfield  {author} {\bibinfo {author} {\bibfnamefont {K.}~\bibnamefont
  {Takubo}}, \bibinfo {author} {\bibfnamefont {R.}~\bibnamefont {Kubota}},
  \bibinfo {author} {\bibfnamefont {T.}~\bibnamefont {Suzuki}}, \bibinfo
  {author} {\bibfnamefont {T.}~\bibnamefont {Kanzaki}}, \bibinfo {author}
  {\bibfnamefont {S.}~\bibnamefont {Miyahara}}, \bibinfo {author}
  {\bibfnamefont {N.}~\bibnamefont {Furukawa}}, \ and\ \bibinfo {author}
  {\bibfnamefont {T.}~\bibnamefont {Katsufuji}},\ }\href {\doibase
  10.1103/PhysRevB.84.094406} {\bibfield  {journal} {\bibinfo  {journal} {Phys.
  Rev. B}\ }\textbf {\bibinfo {volume} {84}},\ \bibinfo {pages} {094406}
  (\bibinfo {year} {2011})}\BibitemShut {NoStop}%
\bibitem [{\citenamefont {Gleason}\ \emph {et~al.}(2014)\citenamefont
  {Gleason}, \citenamefont {Byrum}, \citenamefont {Gim}, \citenamefont
  {Thaler}, \citenamefont {Abbamonte}, \citenamefont {MacDougall},
  \citenamefont {Martin}, \citenamefont {Zhou},\ and\ \citenamefont
  {Cooper}}]{Gleason2014}%
  \BibitemOpen
  \bibfield  {author} {\bibinfo {author} {\bibfnamefont {S.~L.}\ \bibnamefont
  {Gleason}}, \bibinfo {author} {\bibfnamefont {T.}~\bibnamefont {Byrum}},
  \bibinfo {author} {\bibfnamefont {Y.}~\bibnamefont {Gim}}, \bibinfo {author}
  {\bibfnamefont {A.}~\bibnamefont {Thaler}}, \bibinfo {author} {\bibfnamefont
  {P.}~\bibnamefont {Abbamonte}}, \bibinfo {author} {\bibfnamefont {G.~J.}\
  \bibnamefont {MacDougall}}, \bibinfo {author} {\bibfnamefont {L.~W.}\
  \bibnamefont {Martin}}, \bibinfo {author} {\bibfnamefont {H.~D.}\
  \bibnamefont {Zhou}}, \ and\ \bibinfo {author} {\bibfnamefont {S.~L.}\
  \bibnamefont {Cooper}},\ }\href {\doibase 10.1103/PhysRevB.89.134402}
  {\bibfield  {journal} {\bibinfo  {journal} {Physical Review B}\ }\textbf
  {\bibinfo {volume} {89}},\ \bibinfo {pages} {134402} (\bibinfo {year}
  {2014})}\BibitemShut {NoStop}%
\bibitem [{CDO()}]{CDOremnant}%
  \BibitemOpen
  \href@noop {} {}\bibinfo {note} {The peak at 54~cm$^{-1}$ represents an
  excitation of the CDO phase with a relatively large cross-section, as seen in
  Fig.~\ref{doubled_polarization}. A nonzero peak intensity at 54~cm$^{-1}$ in
  the FCO phase is due to the presence of a small, remnant CDO phase fraction,
  as evidenced by a nonzero peak intensity at 295~cm$^{-1}$.}\BibitemShut
  {Stop}%
\bibitem [{FCO()}]{FCOremnant}%
  \BibitemOpen
  \href@noop {} {}\bibinfo {note} {The weak peak at 289~cm$^{-1}$ indicates a
  minority FCO phase is present.}\BibitemShut {Stop}%
\bibitem [{Kim()}]{Kimfootnote}%
  \BibitemOpen
  \href@noop {} {}\bibinfo {note} {See Fig.~3(a) in ref.~[\onlinecite{Kim2010}]
  and Fig.~8(a) in ref.~[\onlinecite{Kim2011a}].}\BibitemShut {Stop}%
\bibitem [{\citenamefont {Guillou}\ \emph {et~al.}(2011)\citenamefont
  {Guillou}, \citenamefont {Thota}, \citenamefont {Prellier}, \citenamefont
  {Kumar},\ and\ \citenamefont {Hardy}}]{Guillou2011}%
  \BibitemOpen
  \bibfield  {author} {\bibinfo {author} {\bibfnamefont {F.}~\bibnamefont
  {Guillou}}, \bibinfo {author} {\bibfnamefont {S.}~\bibnamefont {Thota}},
  \bibinfo {author} {\bibfnamefont {W.}~\bibnamefont {Prellier}}, \bibinfo
  {author} {\bibfnamefont {J.}~\bibnamefont {Kumar}}, \ and\ \bibinfo {author}
  {\bibfnamefont {V.}~\bibnamefont {Hardy}},\ }\href {\doibase
  10.1103/PhysRevB.83.094423} {\bibfield  {journal} {\bibinfo  {journal} {Phys.
  Rev. B}\ }\textbf {\bibinfo {volume} {83}},\ \bibinfo {pages} {094423}
  (\bibinfo {year} {2011})}\BibitemShut {NoStop}%
\end{thebibliography}

%merlin.mbs apsrev4-1.bst 2010-07-25 4.21a (PWD, AO, DPC) hacked
%Control: key (0)
%Control: author (8) initials jnrlst
%Control: editor formatted (1) identically to author
%Control: production of article title (-1) disabled
%Control: page (0) single
%Control: year (1) truncated
%Control: production of eprint (0) enabled
%

\end{document}